\documentclass[preprint]{aastex62}
\DeclareGraphicsExtensions{.eps,.pdf,.png,.jpg}

\received{}
\revised{}
\accepted{}
\submitjournal{AJ}

\shorttitle{Red Supergiant Mass Loss History}
\shortauthors{Humphreys et al.}

\begin{document}

\title{Exploring the Mass Loss Histories of the Red Supergiants\footnote{Based on observations obtained with (1) the NASA/DLR Stratospheric Observatory(SOFIA). SOFIA is jointly operated by the Universities Space Research Association, Inc.(USRA) under NASA contract NAS2-97001 and the Deutsches SOFIA Institut(DSI) under DLR contract 50OK 0901 to the University of Stuttgart, and (2) LMIRCam/LBTI on the Large Binocular Telescope(LBT) an international collaboration among institutions in the United States, Italy, and Germany.LBT Corporation partners include The University of Arizona on behalf of the Arizona university system; Istituto Nazionale di Astrofisica, Italy; LBTBeteiligungsgesellschaft, Germany, representing the Max-Planck Society, the Astrophysical Institute Potsdam, and Heidelberg University; The Ohio State University, and The Research Corporation, on behalf of The University of Notre Dame, University of Minnesota, and University of Virginia.} }

\correspondingauthor{Roberta Humphreys}
\email{roberta@umn.edu}

\author[0000-0003-1720-9807]{Roberta M. Humphreys}
\affiliation{Minnesota Institute for Astrophysics,  
University of Minnesota,
Minneapolis, MN 55455, USA}
\nocollaboration

\author[0000-0002-6542-2920]{Greta Helmel}
\affiliation{Minnesota Institute for Astrophysics,
University of Minnesota,
Minneapolis, MN 55455, USA}
\nocollaboration

\author[0000-0002-8716-6980]{Terry J. Jones}
\affiliation{Minnesota Institute for Astrophysics,
University of Minnesota,
Minneapolis, MN 55455, USA}
\nocollaboration

\author[0000-0002-1913-2682]{Michael S. Gordon}
\affiliation{USRA-SOFIA Science Center, NASA Ames Research Center,
Moffett Field, CA 94035, USA }
\nocollaboration

\begin{abstract}
	We report mid- to far-infrared imaging and photomety from 7 to 37$\mu$m  with SOFIA/FORCAST and 2$\mu$m adaptive optics imaging with LBTI/LMIRCam of a large sample of red supergiants (RSGs) in four Galactic clusters; RSGC1, RSGC2=Stephenson 2, RSGC3, and NGC 7419.  The red supergiants in these clusters cover their expected range in luminosity and initial mass from $\approx$ 9 to more than 25 M$_{\odot}$.  The population includes examples of very late-type RSGs such as MY Cep which may be near the end of the RSG stage, high mass losing maser sources, yellow hypergiants and post-RSG candidates.  Many of the stars and almost all of the most luminous have spectral energy distributions (SEDs) with extended infrared excess radiation at the longest wavelengths. To best model their SEDs we use DUSTY with a variable  radial density distribution function to estimate their mass loss rates.  Our $\dot{M}$ -- luminosity relation for  42 RSGs basically  follows the classical de Jager curve, but at luminosities below 10$^{5}$ L$_{\odot}$  we find a significant population of red supergiants with $\dot{M}$ below the de Jager relation. At luminosities above 10$^{5}$ L$_{\odot}$   there is a rapid transition to  higher mass loss rates that approximates and overlaps the de Jager curve.  We recommend that instead of using a linear relation  or single curve,  the empirical $\dot{M}$ -- luminosity relation is better represented by a broad band. Interestingly, the transition to much higher $\dot{M}$ at about  10$^{5}$ L$_{\odot}$ corresponds approximately to an initial mass of 18 --20 M$_{\odot}$ which is close to the upper limit for RSGs becoming Type II SNe.
\end{abstract}


\section{Introduction} \label{sec:intro}

For many years, the standard view of  massive star
evolution progressed from blue supergiant or O star to red supergiant to 
terminal explosion as a supernova. We now know that the evolution and 
eventual fate of massive stars depends not only on their initial mass, but also on mass loss 
and their mass loss histories. It has been acknowledged for some time that  
the most massive stars do not evolve to the red supergiant region
due to post-main sequence  enhanced mass loss. The observations
suggest an upper limit to the initial mass of  stars that become red
supergiants, corresponding to $\approx$ 40M$_{\odot}$. Thus, the majority of    
massive stars, those between  9  and 30 or 40 M$_{\odot}$,  will pass 
through the red supergiant stage, an important end product of stellar evolution.  

\vspace{2mm} 

\citet{Smartt09,Smartt15} has  identified what he calls ``the red supergiant problem,'' the lack of Type II-P and Type II-L SNe progenitors, usually 
considered to be red supergiants (RSGs), with initial masses greater than 18~M$_{\odot}$.  
Thus, the most luminous and most massive RSGs (18 -- 30 M$_{\odot}$) would 
presumably end their lives some other way.
The RSG stage is a high mass-losing stage. To
what extent mass loss and their mass loss histories can affect the
terminal state of the RSGs and their warmer counterparts is an open question. 
Even though the mass loss mechanism for RSGs is still debated, we can measure
their mass loss from the thermal infrared emission from their circumstellar dust.  Depending on their mass loss, they  
may evolve back to warmer temperatures becoming warm hypergiants, possibly
LBVs, and even WR stars before the terminal explosion or possibly collapse 
directly to black holes. How  they end their lives is pivotal for  stellar 
evolution and the enrichment of the interstellar medium.

\vspace{2mm}

In a series of papers by \citet{Shenoy13,Shenoy15,Shenoy16} and 
\citet{Gordon18},  we have demonstrated that we can map the geometry of the 
circumstellar ejecta and 
reconstruct the mass loss histories of the evolved warm and cool hypergiants  
over the past 1000 years using a combination of high spatial resolution 
near- and mid-infrared imaging and polarimetry over a wide range of 
wavelengths. In combination with SOFIA/FORCAST imaging
 at much longer wavelengths, we can extend their  mass-loss histories to  
 several thousand years. In \citet{Shenoy16}, hereafter Paper I, we  mapped the 
 cold dust in the mid- to far-infrared and discussed the  mass loss histories 
 of the famous hypergiants, VY CMa, IRC~+10420 and 
 $\rho$ Cas,  with some of the highest known mass loss rates. In our 
 second paper, \citet{Gordon18}, hereafter Paper II, we extended our 
 sample with 
 SOFIA/FORCAST observations of three additional strong infrared  and maser 
 sources,
 NML Cyg, VX Sgr and S Per, plus two normal or more typical RSGs, T Per and 
 RS Per. 
 
\vspace{2mm} 
 
Our two previous studies concentrated on individual stars, most with well-known strong infrared emission from dust and high mass loss rates. The extended circumstellar emission of the hypergiants, VY CMa and IRC~10420, revealed their
complex mass loss histories, and many of the most luminous RSGs showed evidence for variable mass loss over the past thousand years or so, hinting at episodic 
high mass loss events.  Although, it is well known that the RSG  mass loss rate  increases with luminosity, it is not known if the stars evolve during the RSG 
stage becoming cooler, with more extended envelopes and higher mass-loss before the terminal state. To examine this question about red supergiant 
evolution, we have obtained mid- to far-infrared imaging with SOFIA/FORCAST and near-infrared adaptive optics imaging with LBT/LMIRCam of coeval 
samples of RSGs in  Galactic clusters.

\vspace{2mm} 

We have selected  four clusters with numerous RSGs.  {\it RSGC1} \citep{Figer,Davies08}, 
{\it RSGC2} = Stephenson2 \citep{Davies07,Stephenson} and {\it RSGC3} \citep{Alexander,Clark} are massive clusters with remarkably large populations of RSGs and 
evolved stars.  The fourth cluster is {\it NGC 7419} \citep{Marco}  with
      five RSGs, one with a very late spectral type (MY Cep, M7 I) and high luminosity, plus a large number of confirmed B and Be-type members.  The red supergiants 
      in the clusters  are presumably coeval with known distances and the same composition.  They are thus an ideal population for comparing the stars' mass loss histories with their derived parameters such as such as luminosity, spectral type, 
      and initial mass at a fixed age and in a known environment. 

\vspace{2mm}

In the next section we describe the observations and data reduction 
with SOFIA/FORCAST and LBT/LMIRCam. In \S {3} we briefly discuss our use of DUSTY \citep{Ivezic} and  
the determination of the mass loss rates. In \S {4} we describe the results for 
the four clusters. We discuss the mass-loss rate-luminosity relation for red suergiants and summarize our conclusions with respect to their  evolutionary 
state \S {5}, and in a brief final section we summarize our recommendations with respect to their mass loss rates.

\section{Observations and Data Reduction} \label{sec:obs}

\subsection{SOFIA/FORCAST:Mid- to Far- IR imaging (5 -- 37 $\mu$m) }

The four clusters were observed with SOFIA/FORCAST during Cycles 5 and 6 in two independent programs; 05-0064 PI: Smith  for RSGC1 and RSGC3 and 06-0089 PI: Humphreys for RSGC2 and NGC 7419. The observations are summarized in Table 1.

\begin{deluxetable*}{llll}[htbp]   
\tabletypesize{\footnotesize}
\tablenum{1}
\tablecaption{SOFIA/FORCAST Observations}
\tablewidth{0pt}
\tablehead{
\colhead{Cluster} & 
\colhead{Date} & 
\colhead{Filters} & 
\colhead{Program}  
} 
\startdata 
RSGC1  & 2017 Aug 02    &  F056,F077,F111,F253,F315 &  05-0064 \\
RSGC2 &  2018 Aug 29, 31 & F077,F111,F197,F253,F315,F371 &  06-0089 \\
RSGC3 &  2017 Aug 03, 07  & F056,F077,F111,F253,F315 &  05-0064 \\ 
NGC 7419 & 2018 Aug 28   & F111, F315, F371\tablenotemark{a} &  06-0089 
\enddata
\tablenotetext{a}{Reduced usable data (level 3 or 4) were not available for filters F077,F197,F253 for NGC 7419.}
\end{deluxetable*}

FORCAST is a dual-channel mid-IR imager covering the 5 to 40 $\mu$m range. Each channel uses a 256 $\times$ 256 pixel blocked-impurity-band (BiB) array and 
provides a distortion-corrected 3$\arcmin$.2 $\times$ 3$\arcmin$.2 field 
of view
with a scale of 0$\farcs$768 per pixel. FORCAST achieves near-diffraction limited imaging, with a PSF FWHM of $\sim 3\farcs7$ in the longest filters. 

Due to the complexities of most of the cluster fields with multiple sources
 a large chop throw was required.  To mitigate coma effects, we used the 
 asymmetric chop C2NC2 mode. With a much sparser population 
 and lower background contamination,  NGC 7419 was observed  in NMC mode.
  In NMC mode, the chop is symmetric about the optical axis of the telescope with one of the two chop positions centered on the target, and the nod is anti-parallel to this chop throw.  C2NC2 is useful for large extended objects or, in the case of these RSG clusters, targets within crowded fields. In this mode, the chop throw is asymmetric, such that one chop position is centered on the optical axis (and the target) while the second (sky) position is off-axis. Rather than nodding, the telescope then slews to an offset position free of sources or significant background and the same chop pattern is repeated. Both of these SOFIA/FORCAST observing strategies offer the same imaging sensitivity and data quality, with NMC mode being much more efficient in terms of on-source telescope time.

We use the level 3 or 4 flux-calibrated images from the SOFA/FORCAST pipeline  
which
corrects for bad pixels, removes sky and telescope background emission 
and coadds the aligned and merged images.  
Aperture photometry was then measured  using the open-source Astropy (Astropy Collaboration et al. 2013)  photutils  package with 10$\arcsec$ apertures. A larger 15$\arcsec$ aperture was used for selected stars such as MY Cep to include
possible emission in their extended profiles.  The photometry for each cluster is reported and discussed in \S {4}. 
The images obtained in Cycle 6 in 2018 were adversely affected by the
degradation of the FORCAST entrance window. This limited the
usefulness of some filters, especially the shorter wavelength filters for
NGC 7419. 

The new FORCAST fluxes for the measured stars are given in the Appendix. The reported errors are from the measured uncertainty in the sky background annulus.  

\subsection{Other Infrared Sources: Near- to Mid- IR Photometry (2 -- 20 $\mu$m)}  

One of our goals with our SOFIA/FORCAST observations was to extend the available photometry to longer wavelengths to search for the presence of cooler 
circumstellar dust for a more complete picture of their mass loss histories.   
We combine our new long wavelength  photometry from FORCAST with 
published fluxes  from existing infrared surveys such as 2MASS \citep{2MASS}, {\it Spitzer}/IRAC Glimpse survey \citep{Benjamin}, WISE \citep{Wright}, MSX \citep{Egan}, and AKARI \citep{Murakami} to produce 
the SEDs 
for the cluster members discussed here.  Not all of the cluster stars were in the FORCAST field of view, but are included in our analysis using data 
from the above sources. Color-coded identification for the different sources for the photometry is used   on the selected SEDs shown in the figures throughout this paper.  

\subsection{LBT/LMIRCam: Adaptive Optics Near-IR Imaging (2$\mu$m)} 

We also obtained high spatial resolution images of several of our program stars with LMIRCam \citep{skru10} with the LBTI \citep{hinz16} using a single 8.4 m primary mirror of the Large Binocular Telescope (LBT).  The focal plane scale was $0.0106\arcsec$ per pixel and all images were made in the Ks $(2.16 ~\micron)$ filter. The observing 
log is listed in Table 2 where the star used for the night's point spread function (PSF) is given in column 2. 
The AO secondary was operating in natural guide star mode for all observations. The telescope guides or ``locks-on'' to the target which must be bright enough
in the visual-red detector for this purpose. Consequently, many of the faint stars in these clusters, especially RSGC1, were not observed. 

All of the PSF stars were observed with combinations of individual frame time and apparent brightness that resulted in images that did not contain saturated pixels. All of the program stars were overexposed to a varying degree to bring up the brightness at the target radii. The program stars divide roughly into two groups, those that were moderately overexposed and have usable data starting at $\sim 0\farcs 1$ (group 1) and a group that were more heavily 
overexposed and have usable data starting at $\sim 0\farcs 5$ (group 2). 
these groupings are indicated in column 4 of Table 2.

An azimuthally averaged profile for the PSF star HD 216721 is shown in Figure 1 where the vertical axis is the ratio of the surface brightness in ADUs-per-pixel to the total ADUs for the stellar image.  Clearly visible are the first three Airy rings at 0.09, 0.15 and $0.2$ arcseconds. This star and GSC 04014-0230 were used as the PSF standard for stars in group 1. The PSF star observations for group 1 had very stable FWHM values, but there was some variation in the wings beyond $0\farcs3$, outside the region we are using.

The azimuthally averaged profile for GSPC S875-C is also shown in Figure 1, where the vertical axis is the same as for HD 216721. For this PSF star, only the region between $0\farcs 5$ and $1\farcs0$ was used for the stars in group 2. For group 1, we will be investigating scattered light in the mass loss wind at radial distances of $\sim 0\farcs1$, about 300 AU at a distance of 3 kpc. For group 2, we will be investigating scattered light in the mass loss wind at radial distances of $\sim 0\farcs7$, about 4200 AU at a distance of 6 kpc.   As a check on our PSF for group 2, direct comparison between radial profiles of stars in the frame and between frames in this group were compared. Only RSGC-02 showed any clear difference (an excess) from the other stars, or the PSF star.

\begin{deluxetable}{cccc}[htbp]   
\tablecaption{LMIRCam Observations at $2.16~\micron$}
\tablenum{2}
\tablewidth{0pt}
\tablehead{ \colhead{Target Star} & \colhead{PSF Star} & \colhead{Date} & \colhead{Group} }
\startdata
N7419-B950(MY Cep)  & HD 216721 & 2016 Oct 12 & 1 \\
N7419-B139 & HD 216721 & 2016 Oct 12 & 1 \\
RSGC1-04 & GSPC S875-C & 2017 Apr 10 & 2 \\
RSGC2-02  & GSPC S875-C & 2017 Apr 10 & 2 \\
RSGC2-03  & GSPC S875-C & 2017 May 20 & 2 \\
RSGC2-08  & GSPC S875-C & 2017 May 20 & 2 \\
RSGC2-06  & GSPC S875-C & 2017 May 20 & 2 \\
M1-s04\tablenotemark{a}  & GSPC S875-C & 2017 May 21 & 2 \\
M1-s03\tablenotemark{a}  & GSPC S875-C & 2017 May 21 & 2 \\
N7419-B696  & GSC 04014-0230 & 2017 Oct 03 & 1 \\
\enddata
	\tablenotetext{a}{These two RSGs from Masgomas-1 \citep{Ramirez} were added to the LMIRCam program because they were bright enough for the AO guiding system.} 
\end{deluxetable}

\begin{figure}[h]  
\epsscale{1.0}
\plottwo{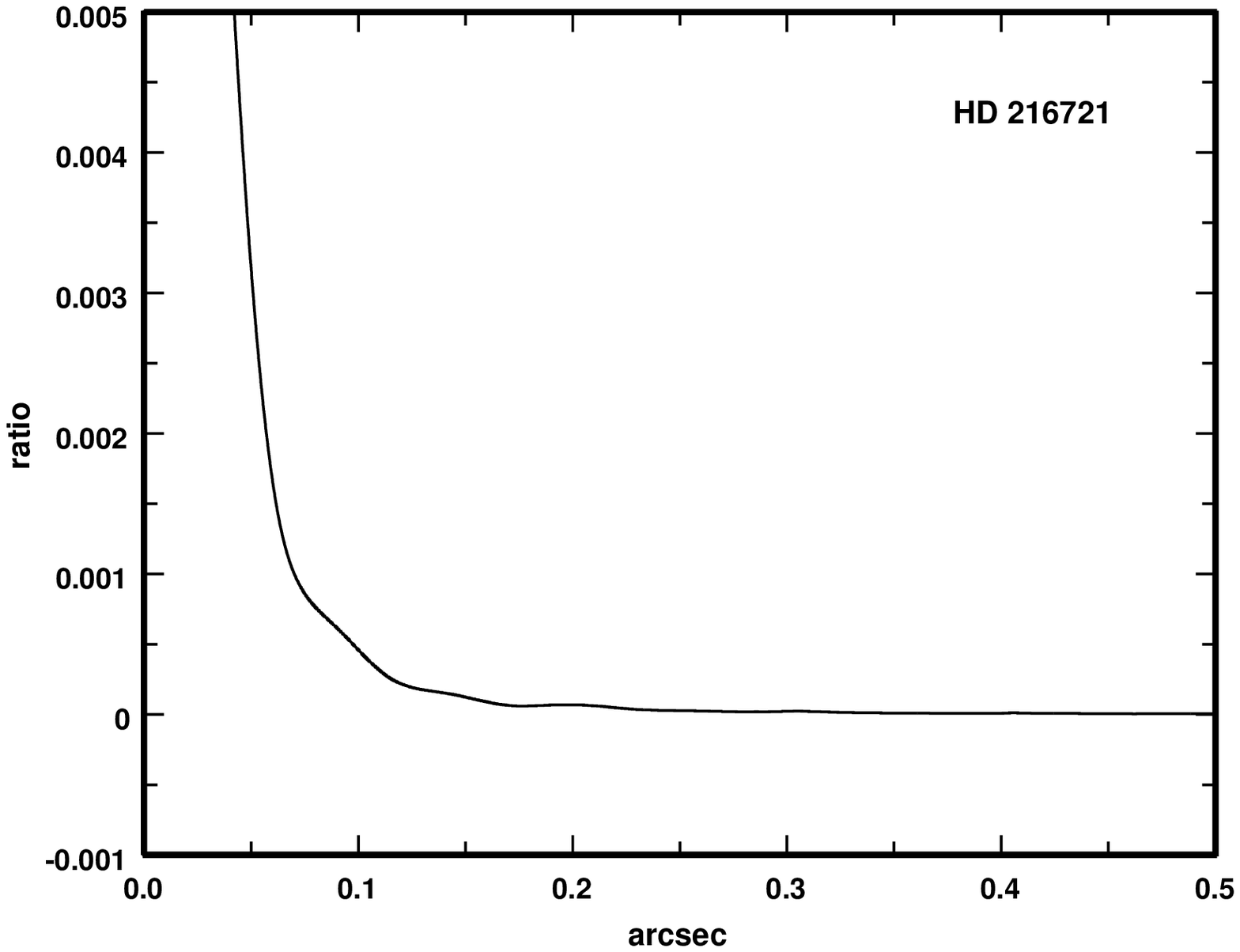}{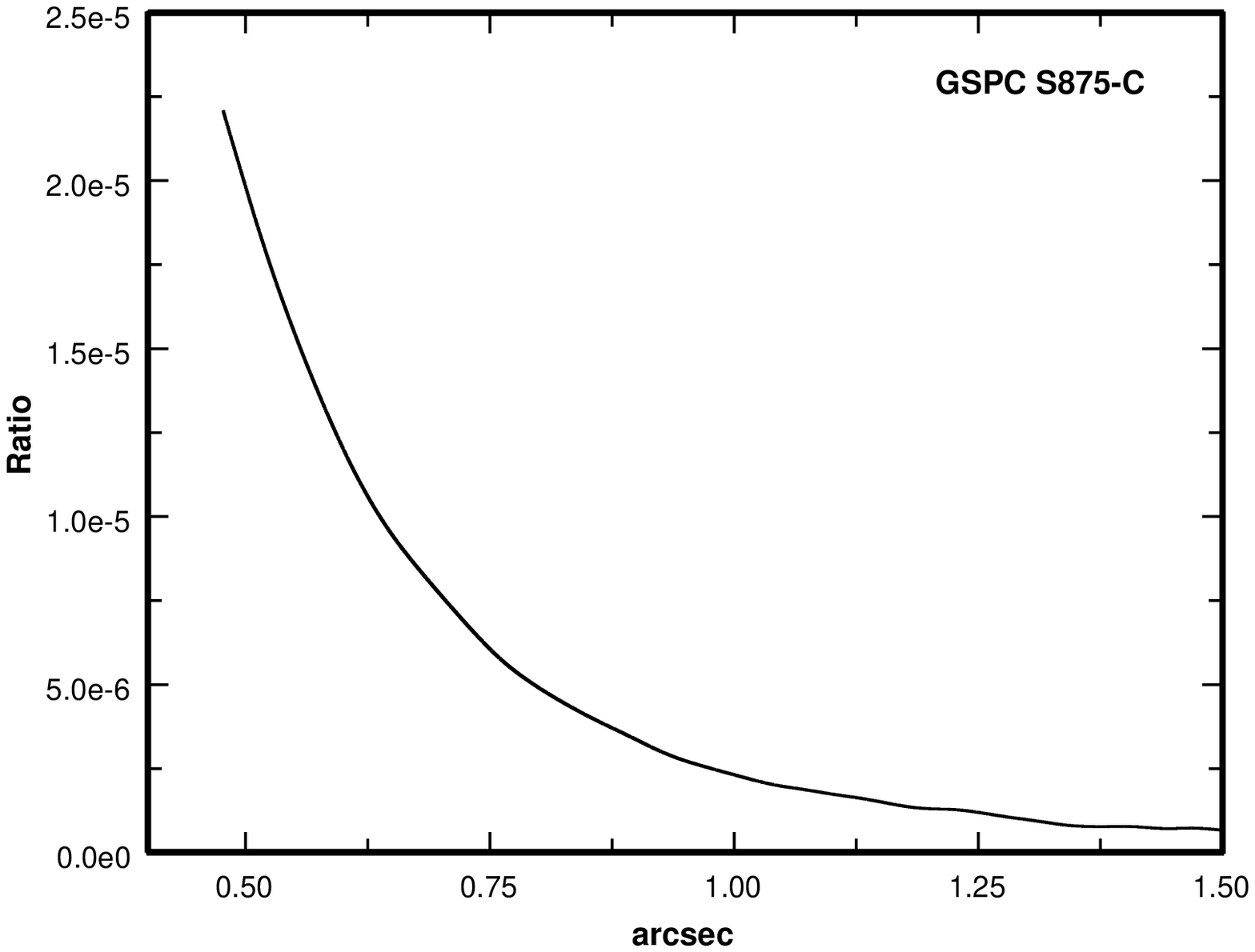}
\caption{Left: Azimuthally averaged radial profile of the PSF star HD 216721. Vertical
 units are the ratio of the surface brightness (flux per pixel) to the total flux in the stellar image. Right:  Azimuthally averaged radial profile of the PSF star GSPC S875-C. Vertical units are the ratio of the surface brightness (flux per pixel) to the total flux in the stellar image.}
\end{figure}

\begin{figure}[h!]  
	\plottwo{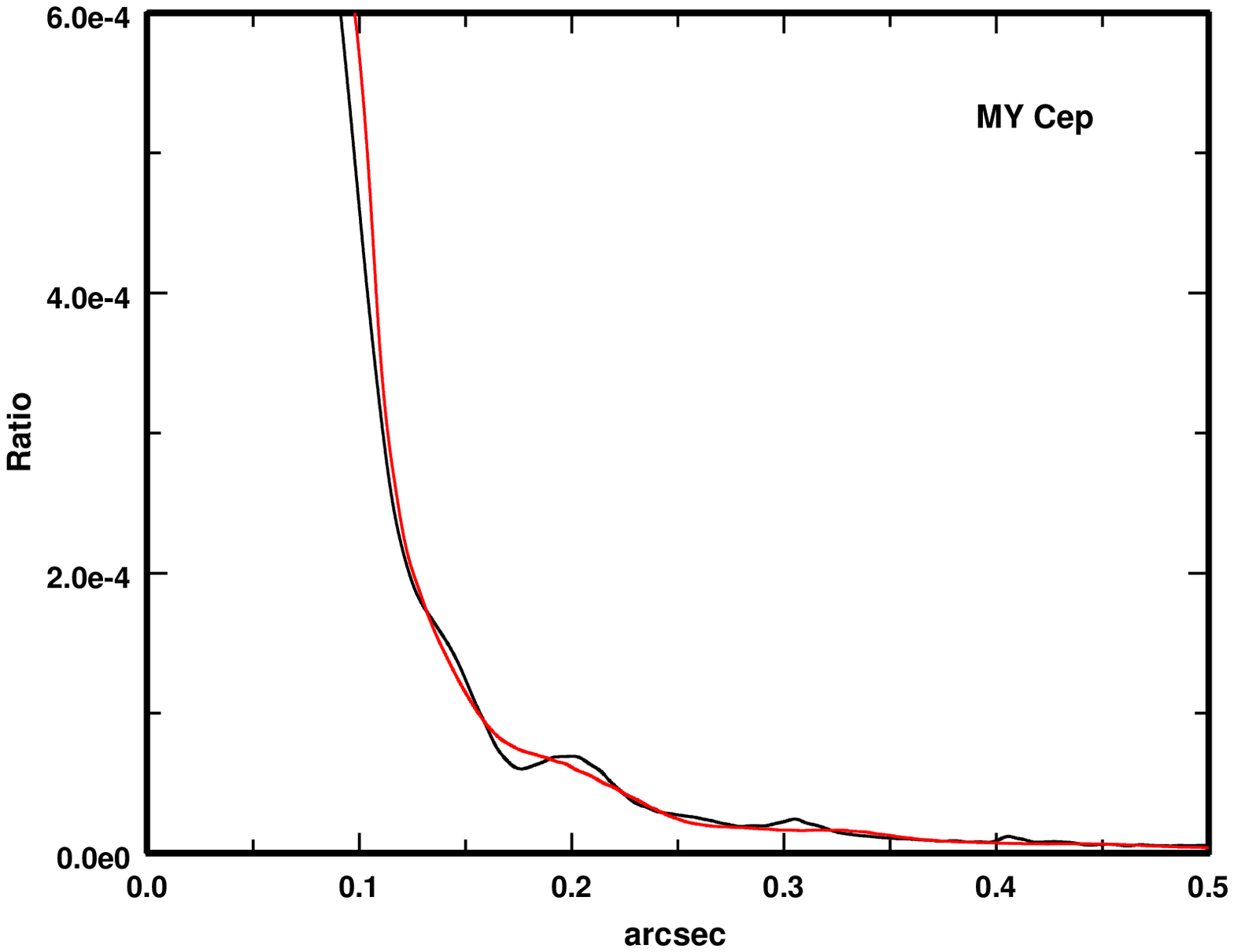}{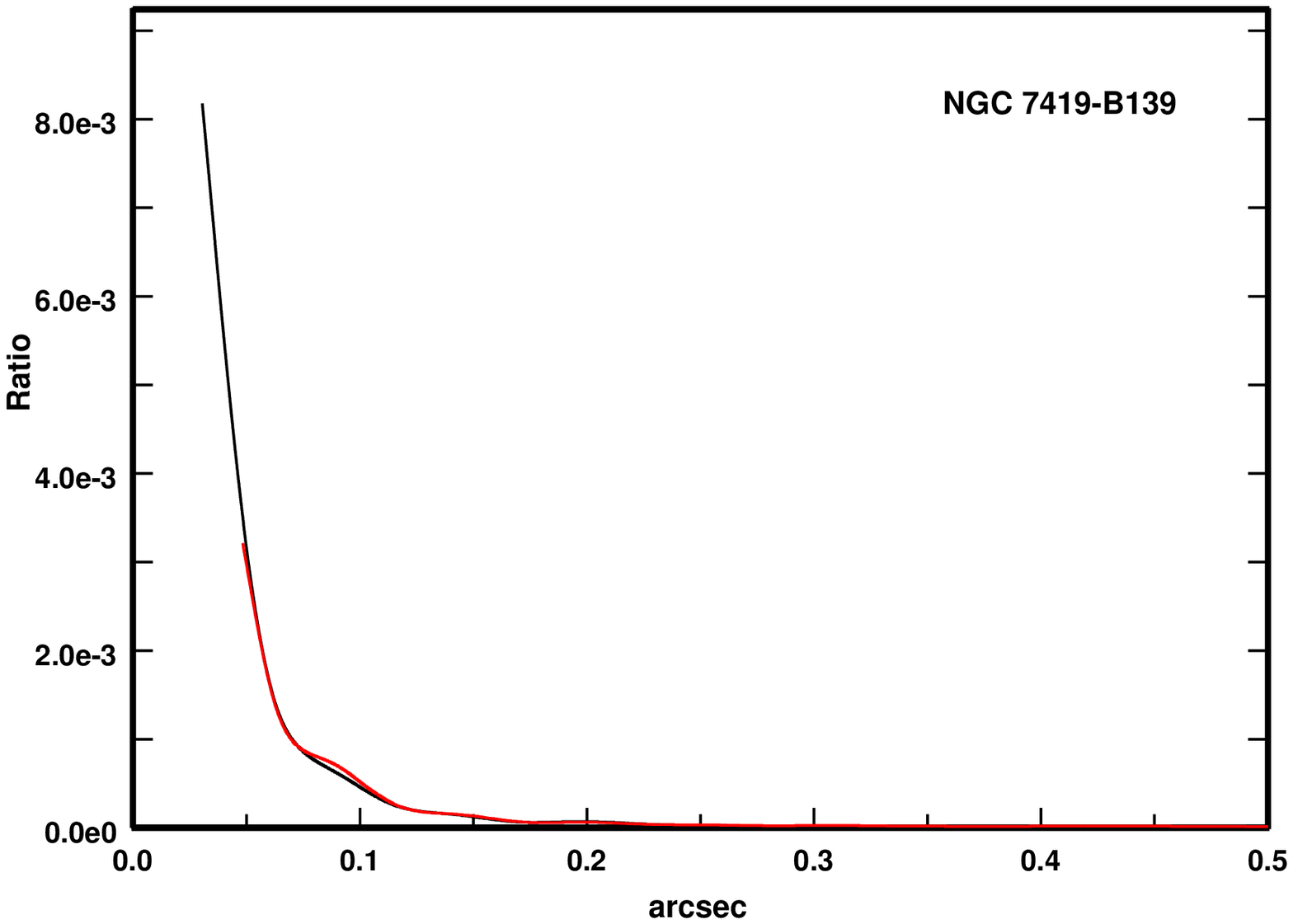}
	\caption{\label{fig:B139}The averaged radial profile for MY Cep (B950) compared with a PSF standard. The vertical axis is the contrast ratio in flux per pixel divided by the total flux from the star. MY Cep is in group 1, and shows extended emission within a radius of $0.1\arcsec$, about 300AU. Right: Averaged radial profile for NGC 7419-B139 compared to a PSF standard. This star also in group 1 did not show any measurable extended emission.}
\end{figure}

\begin{figure}[h!] 
	\plottwo{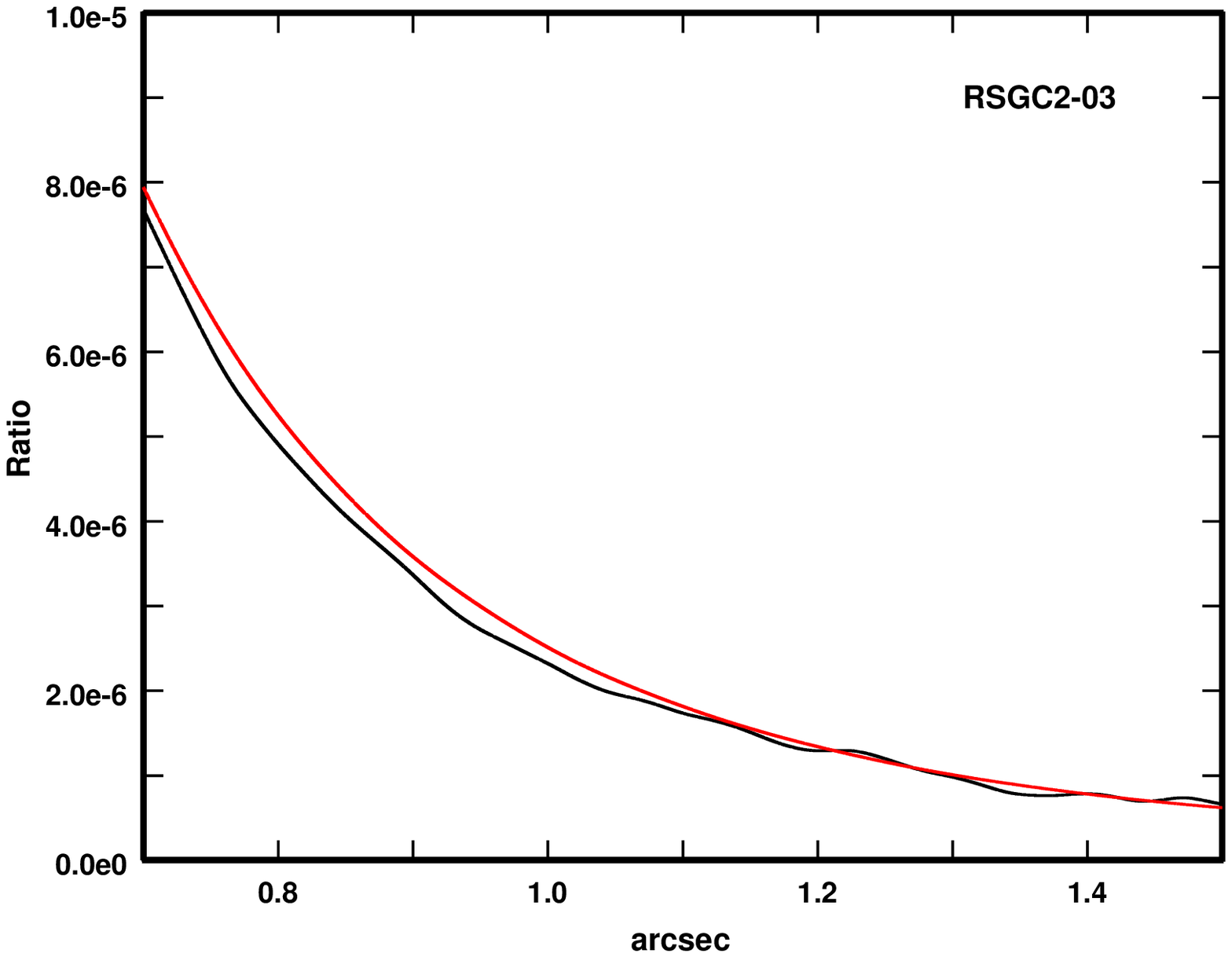}{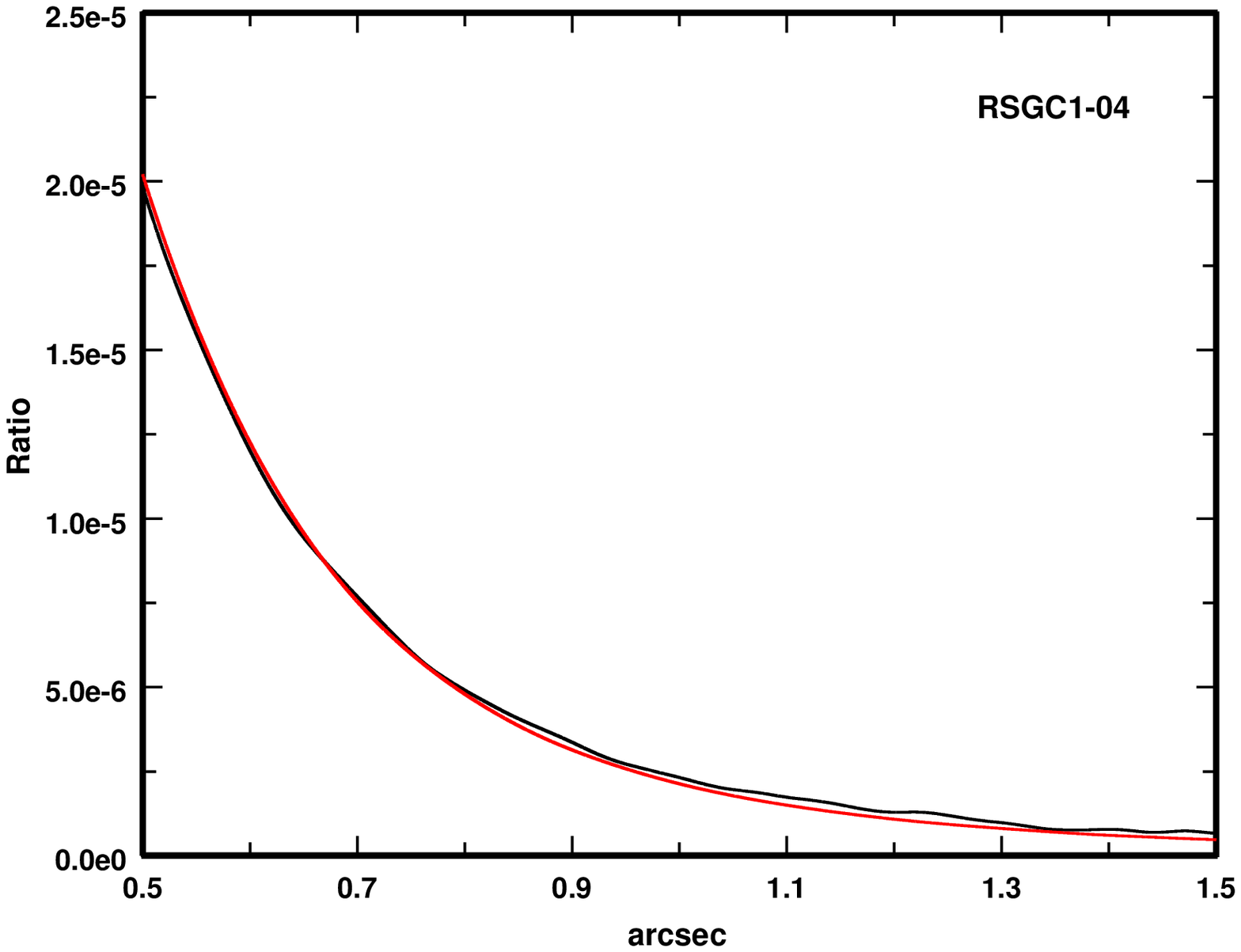}
\caption{\label{fig:S03}Left: Averaged radial profile for RSGC2-03 compared with a the PSF standard. The vertical axis is the contrast ratio in flux per pixel divided by the total flux from the star. This star is in group 2, and shows measurable extended emission. The stellar profile begins to enter the non-linear regime at radii less than $0.6\arcsec$. Right: Averaged radial profile for RSGC1-04 compared to a PSF standard. It  shows no measurable extended emission.}
\end{figure}

The azimuthally averaged radial profile for each star observed with LMIRCam was computed and compared to the mean radial profile of the PSF standard observed on the same night (see Table 2). No 2D information on the sky was measured, only the mean radial profile. Four examples are shown in Figures \ref{fig:B139} and \ref{fig:S03}, one each from group 1 and 2 described in \S {2.3} that had measurable extended emission, and one each in the two groups that did not. In all cases the vertical axis is the ratio of the amount of flux in a single pixel to the total flux from the star in the image. For highly saturated stars, the total flux was estimated by matching the stellar profile to the PSF at large radii ($1.1-1.5\arcsec$, and using the PSF profile (which is unsaturated) to extrapolate to the total flux. Since errors in total flux only slightly influence the PSF subtraction, we did not make a comparison of total flux with 2MASS photometry. The major uncertainty in measuring net emission above the PSF profile is not raw signal-to-noise, as there is plenty of signal. Rather, systematic effects that can influence the width of the stellar profile, which dominates the systematic effects for group 1,  and fitting the PSF at large radii, which dominates the systematic effects for group 2, are the primary sources of our estimate of the uncertainty. 

\section{DUSTY Modeling and Mass Loss} 

To estimate the mass-loss rates and determine the density distribution of the gas and dust, we use the DUSTY radiative-transfer code \citep{Ivezic} to model the observed 
SEDs similar to our Papers I and II. DUSTY solves the one-dimensional radiative transfer 
equation for a spherically symmetric dust distribution.  The input includes the optical properties of the dust, the chemistry, size of the 
 grains, and a dust condensation temperature which sets the dust condensation 
 radius, r$_{1}$.  For the dust optical properties, we use the `cool' circumstellar silicates  \citep{Ossen} and assume that the grains follow an MRN size distribution, n(a)$\,\propto$ a$^{-3.5}$da \citep{Mathis}  with a$_{min}$ = 0.005\,$\mu$m and a$_{max}$ = 0.5\,$\mu$m.

We generate a series of models for each star with an adopted stellar temperature
 initially corresponding to its spectral type when available, and a fixed shell extent (1000 r$_{1}$). We chose the Planck curve option in DUSTY to fit the near- and mid -infrared fluxes instead of an atmospheric model. All but two of the red supergiants lack the usable visual photometry for the model fits. Together with the range in the published spectral types and scatter in the fluxes discussed later, we considered the Planck curve fits  adequate for our purposes.

The dust condensation temperature is an important parameter in modeling the wind. Values in the literature vary from 700--1500 K \citep{Cassar} with a range of 700--1000 K for silicate-rich dust shells. In this work,  we adopt a fixed dust condensation temperature of 1000~K for the temperature of the inner shell (T$_{in}$) following \citet{Suh} and consistent with  observations of RSG dust shells \citep{RowanRobinson82}.  We varied the optical depth $\tau_{v}$ of the circumstellar material from  0.01 up to 5, but 
 find that a realistic range for most of the stars in this study is $\tau_{v}$ of  0.1 
 to 1.  We also vary the density distribution function, which in DUSTY is modeled as a power law, $\rho$(r) $\propto$ r$^{-n}$. An index of $n=2$ is for the standard  constant mass-loss rate.  An index less than that indicates a higher 
 mass-loss rate in the past and a decline over the dynamical age.   We initially adopted the constant mass loss density prescription with $\rho$(r) $\propto$ r$^{-2}$ and varied $\tau_{v}$ to get a best fit SED.  However, several 
stars with a large circumstellar excess radiation at long wavelengths required 
a less steep power law.  In this paper we show the DUSTY fits for a constant 
mass-loss rate, $n=2$ and adopt a lower value of $n$ for those stars with large excess radiation at the longest wavelengths. 

\subsection{The Mass-Loss Rate}

If we assume a constant mass-loss rate over time and a uniform expansion 
velocity, the mass-loss rate is the familiar equation $\dot M(t)$ $=$ 4$\pi$r$^{2}$$\rho$(r)\,v$_{exp}$. 
But for stars with a non-constant mass-loss 
rate (n $\neq$ 2), it is necessary to integrate the density distribution over 
the spatial extent of the shell. We discuss the derivation of the form of this equation  based on the input from DUSTY for a non-constant 
mass loss rate  in the Appendix. The mass loss rate for dust in gm s$^{-1}$ is given by:

\begin{eqnarray} \label{eq:1}
\dot M &=& \frac{{{M_{tot}}V}}{{{r_{\max }} - {r_{\min }}}} \sim \frac{{16\pi }}{3}\frac{{n - 1}}{{3 - n}}\sqrt {{a_{\max }}{a_{\min }}} {\rho _d}\frac{\tau }{{{Q_{\rm{eff}}}}}r_{\max }^{2 - n}r_{\min }^{n - 1}v_{exp}
\end{eqnarray}

This is Equation~\ref{eq:A8} in the Appendix. For a constant mass-loss rate, with $n=2$, this becomes:

\begin{eqnarray}
\dot M \sim \frac{{16\pi }}{3}\sqrt {{a_{\max }}{a_{\min }}} {\rho _d}\frac{\tau }{{{Q_{\rm{eff}}}}}{r_{\min }}v_{exp}
\end{eqnarray}

To determine the combined mass loss rate from dust and gas we require a gas to dust ratio (g/d). Most values range from 100 to 200 in the Galaxy with a high of 500 suggested for the Large Magellanic Cloud \citep{vanLoon99}. In our two previous studies \citep{Shenoy16,Gordon18} we adopted a gas to dust ratio of 100. Here we use a g/d ratio of 200 \citep{Decin,Joss} for Galactic RSGs for comparison with  other recent work. We also assume the nominal expansion velocity of 25 km s$^{-1}$. Higher expansion velocities of 35 -- 40 km s$^{-1}$ are measured for the most luminous red supergiants with high mass-loss rates. The impact of these two adopted parameters on our mass-loss rates and conclusions is discussed in \S {5}. 

\subsection{Model Profiles}

DUSTY computes a model radial profile at a chosen wavelength as a function of the inner dust radius $\rm{r}_1$ that can be compared to our LMIRCam observations. Since our LMIRCam images can experience saturation effects at small radii and uncertainties in normalization at larger radii, we chose to measure the excess surface brightness at a specific radii, between these two limitations. The results are given in Table 3. Using the distances in Table 4, we converted the angular distance of these radii to the projected physical distance in AU.

\begin{figure}[h!]  
\plotone{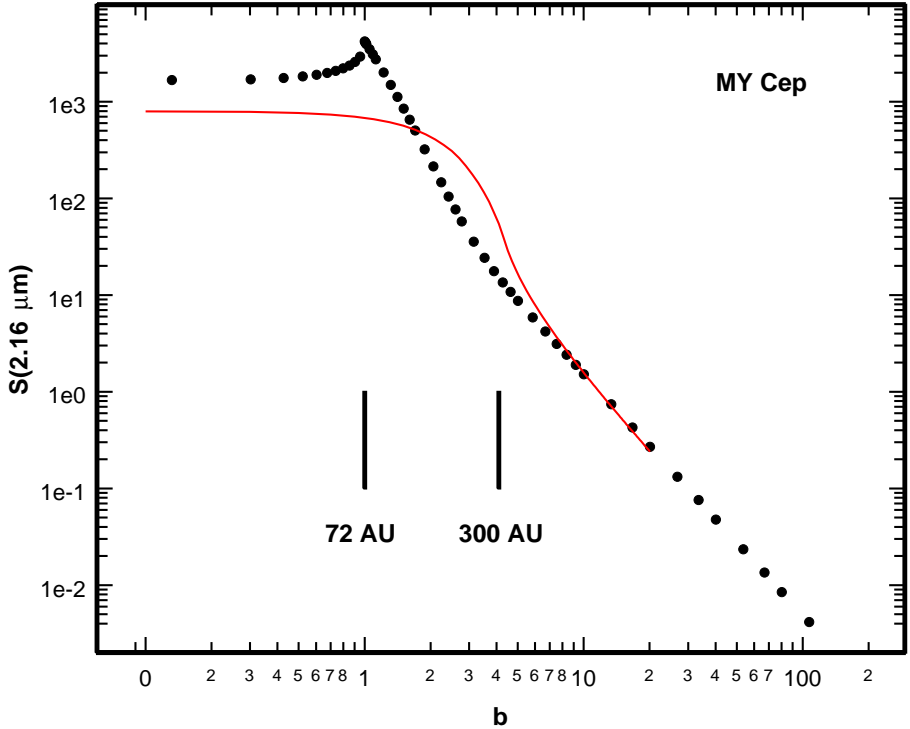}
	\caption{\label{fig:B950stuff} The projected radial profile of the DUSTY model dust shell at $2.16~\micron$ (black dots) as a function of the normalized impact parameter b. At $\rm{r}_1$, b = 1.0. The spike at b = 1 ($\rm{r}_1$ = 72 AU) is the inner edge of the dust shell. The break in the slope at b = 3 is where the primary contribution to the surface brightness changes from thermal to scattered flux. The red line is the expected profile convolved with our PSF. The location of our LMIRCam flux measurement ($0\farcs1$) is at 300 AU.}
\end{figure}

For the three stars in Group 1, the DUSTY profile needs to be convolved with our best PSF for us to determine the model surface brightness for comparison with our observations. DUSTY does output a radial profile that includes the contribution from the central star, but we are interested in just the excess we find in flux above the PSF at a specific radius with our LMIRCam observations. The contribution from the star and the shell are additive in the convolution with the PSF and convolving the dust shell is equivalent to subtracting the PSF from the observed profile for comparison with the observed excess over the PSF for the program star. We take this approach to make the effect of the convolution easier to see, which is important for separating the thermal and scattered emission in the dust. This is illustrated in Figure \ref{fig:B950stuff} for MY Cep where the DUSTY model profile of the dust shell has been convolved with the PSF. Note that the model surface brightness at b = 4.1 (r = 300 AU) is dominated by scattered light, but that blurring by the PSF brings in thermal emission from smaller radii. We can then correct our observed excess flux at that radii for blurring by the PSF. For the 7 stars in Group 2, the excess flux is measured at $0.7\arcsec$, and no convolution is necessary. In this case, a direct comparison of the DUSTY model (if there is one) with the observed excess (if any) can be made.  

We have converted our excess flux (corrected for the PSF) for those stars with a measurable excess into a scattering optical depth using the method outlined in \cite{Shenoy15}. This can be directly compared with the DUSTY model visual optical depth at the same impact parameter b by converting $\tau_{V}$ to $\tau_{sca}$ at $2.16\,\mu$m, using the dust parameters that were inputs to DUSTY. The observed and model scattering optical depths are given in Table 3, where column 3 is the measured scattering optical depth with LMIRCam, corrected for PSF smearing, and column 4 is the model output value from the DUSTY models in the following section. 

\begin{deluxetable}{cccc}
\tablecaption{Extended Scattered Emission}
	\tablenum{3} 
\label{tbl:extended}
\tablewidth{0pt}
	\tablehead{\colhead{Star} & \colhead{radius (AU)} & \colhead{$\tau_{sca}$ $2.2~\micron$} & \colhead{$\tau_{sca}${DUSTY}}}
\startdata
NGC 7419-B950 & 300 & $0.035\pm 0.005$ & 0.08 \\
NGC 7419-B139 & 200 & $<0.02$ & 0.02 \\
NGC 7419-B696 & 200 & $0.015\pm 0.005$ & 0.02 \\
RSGC1-04 & 4500 & $<0.009$ & -\\
RSGC2-02 & 4000 & $<0.009$ & - \\
RSGC2-03 & 4000 & $0.008\pm 0.003$ & 0.0003 \\
RSGC2-06 & 4000 & $<0.009$ & 0.00015 \\
RSGC2-08 & 4000 & $<0.009$ & 0.0002 \\
M1-s04 & 2500 & $<0.009$ & - \\
M1-s03 & 2500 & $<0.009$ & - \\
\enddata
\end{deluxetable}

We were only able to measure excess flux for three of the program stars observed with LMIRCam, and only MY Cep has a solid detection. In the case of MY Cep, the observed optical depth is lower than predicted by the DUSTY model. This is likely due to the fact that DUSTY assumes isotropic scattering by the dust, but at $2.16~\micron$ the dust is higly forward scattering. Using our dust parameters and a wavelength of $2.16\micron$, the ratio of the scattered flux at a $90-180\degr$ scattering angle to the scattered flux at $0\degr$ (forward, away from the star) is calculated to be 0.25. We have not made a detailed integration down the line of sight at b = 4.1 for MY Cep, but given the range of scattering angles at different locations along this line of sight through the dust shell, the ratio between our observed flux and the model flux of 0.44 is reasonable.

For B696, the observed optical depth is comparable to the model prediction, suggesting a slight excess of emission at a 200 AU offset from the star. For group 2 stars, only RSGC2-03 had a measurable excess, ablbeit with S/N = 2.7. The observed scattering optical depth is significantly larger than the model prediction. If the observed value is real, then this star must have had a higher mass-loss episode about 1000 yrs ago, assuming a wind velocity of 20 km/s. Another possibility is that the dust in the outer shell of this star is scattering the diffuse interstellar radiation field within the cluster. None of the other stars in RSGC2 had measurable flux excesses, but their $3\sigma$ upper limits on the observed scattering optical depth are too large for a careful comparison between stars.

 The DUSTY models also provide predicted radial profiles at the longer wavelengths of our SOFIA FORCAST observations. We find that uncertainties in the reproducibility of the PSF between observations of standards and the program stars makes comparison between the FORCAST profiles and the DUSTY models highly uncertain. This is due to both variations in the SOFIA PSF and the large FWHM of $3-4\arcsec$ for the FORCAST images. Convolving the DUSTY profiles with this large PSF results in very small enhancement to the stellar profile due to the presence of the dust shell.

\section{The Clusters}\label{sec:cluster}

The adopted parameters for the four clusters, their distances, and foreground interstellar extinction are summarized in Table 4.  The four clusters represent a significant range in their ages which is reflected in the properties of their
red supergiant populations discussed in this section. For example, RSGC1, the youngest,  has the most luminous RSG members and with the corresponding highest mass loss rates. Thus the three ``RSG'' clusters together provide a good sample for evaluating the evolutionary state of the red supergiants and the role of mass loss. NGC 7419, the closest, is in the middle of the age range, but has an interesting population with the very late-type RSG, MY Cep. 

To estimate the luminosities of the red supergiants, we integrate their 
extinction-corrected SEDs with the adopted temperature from the DUSTY model 
fit to the observations. The greatest source of error in the derived luminosities is the uncertainty in the distance. NGC 7419 has a well-determined distance 
from Gaia parallaxes \citep{DB2019} at 3 kpc (+0.35, -0.29 kpc). This approximately 10\% error in the distance leads to a 20\% uncertainty in luminosity. 
The three ``RSG'' clusters are physically close on the sky, but  much further away at the intersection of the Scutum-Crux spiral arm with the Galactic bar at 
approximately 6 kpc. Given their location they are more likely further away
than significantly closer. In this work, we adopt the published distances with 
the estimated error which is typically $\pm$ 1 kpc corresponding to a
37\% uncertainty in the derived luminosity. Other sources of error in the integrated SED are the adopted temperature, the interstellar extinction, and the fit to the observed fluxes. The combined errors are discussed for each cluster.

The uncertainty in the luminosity also enters into the calculation of the mass loss rate from the DUSTY model; specifically r$_{1}$  in the DUSTY output where this parameter depends on L$^{1/2}$ and on the adopted temperature in  the DUSTY models.  

To correct the observed fluxes for interstellar extinction we adopt the extinction corrections for the 2MASS data \citep{Koorneef} and follow the  extinction law from \citet{Cardelli} for the longer wavelengths measured relative to A$_{K}$.
The adopted correction for each cluster is discussed separately below and summarized in Table 4.

\begin{deluxetable*}{lclllll}[h!]  
\tabletypesize{\footnotesize}
\tablenum{4}
\tablecaption{Cluster Parameters Summary}
\tablewidth{0pt}
\tablehead{
\colhead{Cluster} &
\colhead{N(RSGs)} &
\colhead{Dist(Kpc)} & 
\colhead{IS Ext(mag)} & 
\colhead{Mass M$_{\odot}$ } &
\colhead{Age(Myr)} & 
\colhead{Refs\tablenotemark{a}}   
} 
\startdata 
	NGC7419 & 5 & 3.0 $\pm$ 0.3 & 5.4(A$_{V}$) &  5--10 $\times 10^{3}$ & 14 $\pm$ 2\tablenotemark{c}  & 1, 2, 3  \\  
RSGC1 &  14   &  6.6 $\pm$ 1 & 2.29(A$_{K}$)  &  2--4 $\times 10^{4}$ & 7--12  & 4, 5  \\
RSGC2    & 26 & 5.8 $\pm$ 1 & 1.32(A$_{K}$)\tablenotemark{b} & 4 $\times 10^{4}$ &  12--17 & 6 \\
RSGC3 &  15  &  6 $\pm$ 1 &  1.5(A$_{K}$) &  2--4 $\times 10^{4}$ & 16--20 & 7, 8    
\enddata
	\tablenotetext{a}{References: (1) \citet{Marco}, (2) \citet{BD2018}, (3) \citet{DB2019}, (4) \citet{Figer}, (5) \citet{Davies08},  (6) \citet{Davies07}, (7) \citet{Alexander}, (8) \citet{Clark}}  
\tablenotetext{b}{See text for RSGC2.}
	\tablenotetext{c}{\citet{BD2020} cite ages of 20 $\pm$ 1 for NGC 7419 and 7 $\pm$ 2 for RSGC1.}
\end{deluxetable*}

\subsection{NGC 7419}

At 3kpc, NGC 7419 is relatively nearby  with several published 
studies of its stellar content \citep{Beauchamp94,Subram06,Joshi08}.  It is best 
known for its five red supergiants in one cluster including the very late-type 
MY Cep (M7 --7.5 I ). The most recent study by \citet{Marco}  
includes a comprehensive survey of its Be and B star population as well as the 
red supergiants. 

To assess the total affect on the derived luminosities in addition to the error in the distance, we assume an uncertainty of $\pm$ 100K in the adopted temperatures from the small range in the published spectral types for the RSGs in N7419 or $\approx$ 10\% on the luminosity. The adopted A$_{V}$ is derived from the normal B (non-Be) star members and although no error is quoted we  adopt $\pm$ 0.2 mag or 4\%. The combined error from the residuals to the SED fits for these stars is small corresponding to about 3\% on the derived luminosity.  Combining these errors with the distance, gives an uncertainty of 23\% in luminosity and 14\% in the mass loss rate.  

MY Cep (B950) is especially interesting because it belongs to the rather rare 
late-type and relatively high luminosity RSGs; that is red supergiants with 
spectral types M4 to M5 and later. This group includes stars like S Per, VX Sgr, NML Cyg and VY CMa with high mass loss rates, and extensive circumstellar dust 
and ejecta. All four  are strong maser sources, and MY Cep is also source of 
OH, SiO, and H$_{2}$O maser emission \citep{Verheyen}. 
 MY Cep's spectral energy distribution (SED) in Figure 5 clearly demonstrates the presence of circumstellar dust at the longer wavelengths. We ran DUSTY models 
 with a range of $\tau_{v}$ from 0.1 to 5. 
 The model with $\tau_{v}$  = 1, and T = 3000~K is shown in the 
 figure. It is clear that models with constant mass loss rate do not 
 reproduce the far-infrared flux. Here we adopt $n=1.5$ indicating a higher 
 mass-loss rate in the past.  We integrate the observed best-fit SED fit  
 from $1\,\mu$m to $40\,\mu$m together with a 3000~K Planck curve to fit the shorter wavelengths. The re-radiated flux from the circumstellar dust dominates the SED 
 from 8 to $40\,\mu$m.   
   This gives a total luminosity of 3.1 $\times$ 10$^{5}$ L$_{\odot}$ and a 
   mass- loss rate of 2.3 $\times 10^{-5}$  M$_{\odot}$ yr$^{-1}$ from 
   Equation~\ref{eq:1} (\ref{eq:A8})  with a gas to dust ratio of 200.  

\begin{figure}[h]   
\plottwo{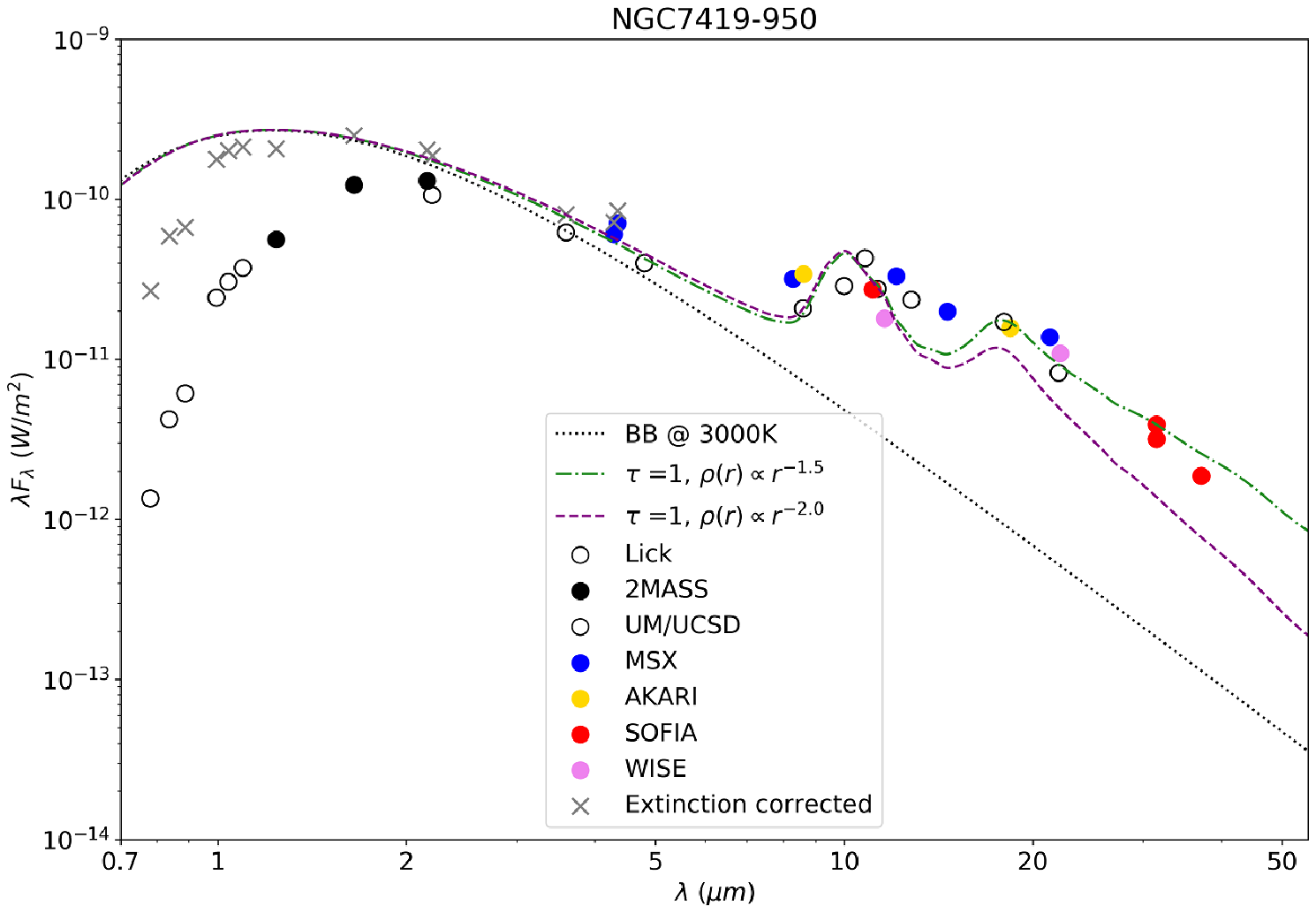}{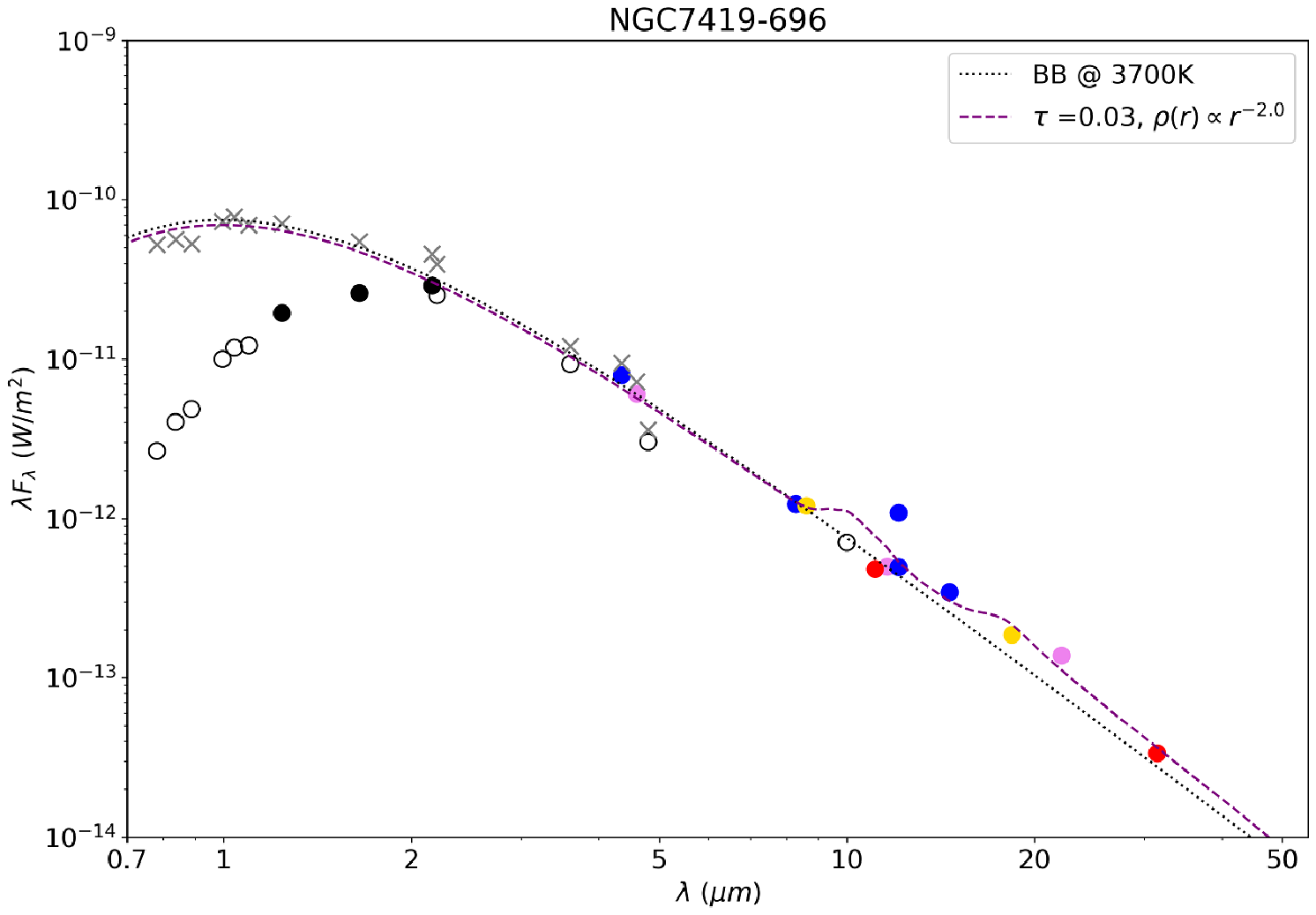}
	\caption{The SEDs for N7419-B950 (MY~Cep), left,  and B696, right, showing the observed (circles) and interstellar extinction corrected (tipped crosses) fluxes from different sources. The symbols are the same in each SED. The open circles are from \citet{Fawley}. A Planck curve for the adopted temperature and the DUSTY model for different power 
	laws are shown. MY~Cep's strong silicate emission feature and longer wavelength fluxes are best fit with a less steep power law ($n=1.5$). B696 has a weak dust emission if any and is fit with a low $\tau_{v}$ and a constant mass-loss rate.}  
\end{figure}

The four other RSGs are significantly less luminous and much earlier in 
spectral type. Their DUSTY models are consistent with a constant mass-loss 
rate with low $\tau_{v}$. The SEDs for B696,  shown in Figure 5, B435, and B139 are  best fit with  a low $\tau_{v}$ and a constant mass-loss rate.  B921 has no long wavelength infrared excess and no evidence for circumstellar dust.  We adopt an upper limit to its mass loss rate, based on the other three stars, in later discussion. Our results for the N7419 RSGs are summarized in Table 5. Note that B139 was not on the FORCAST frame.

\begin{deluxetable*}{llcccll}[htbp]   
\tabletypesize{\footnotesize}
\tablenum{5}
\tablecaption{Model Results for NGC 7419 Red Supergiants\tablenotemark{a}}
\tablewidth{0pt}
\tablehead{
\colhead{Star} & 
\colhead{Sp Type/Temp} & 
\colhead{Power law(r $^{n}$)} & 
\colhead{$\tau_{v}$} &
\colhead{r$_{1}$ (AU)} & 
\colhead{L$_{\odot}$} &
\colhead{M$_{\odot}$ yr$^{-1}$}
}
\startdata 
B950(MY Cep)*  & M7--7.5 I/3000 & 1.5 & 1 & 73 & 3.1 $\pm$ 0.7 $\times$ 10$^{5}$ & 2.3 $\pm$ 0.3 $\times$ 10$^{-5}$ \\
B435*          & M1.5 Iab/3600  & 2.0 & 0.05 & 21 & 2.3 $\pm$ 0.5 $\times$ 10$^{4}$ & 6.4 $\pm$ 0.9 $\times$ 10$^{-8}$ \\
B921*          & M0 Iab/3800  & \nodata & \nodata  & \nodata & 1.4 $\pm$ 0.3 $\times$ 10$^{4}$ & no IR excess \\
	B696*          & M1.5 Iab/3700 & 2.0 & 0.03 & 27 & 3.5 $\pm$ 0.8 $\times$ 10$^{4}$ & 4.9 $\pm$ 0.7 $\times$ 10$^{-8}$ \\
B139         & M1 Iab/3700   & 2.0 & 0.03 & 30 & 4.7 $\pm$ 1.0 $\times$ 10$^{4}$ & 5.5 $\pm$ 0.8 $\times$ 10$^{-8}$ \\
\enddata
\tablenotetext{a}{Stars with new FORCAST fluxes have an asterisk.}
\end{deluxetable*}

\citet{BD2018}  have presented results for DUSTY models for the NGC 7419 RSGs. They treat the DUSTY models somewhat differently, allowing the dust 
condensation temperature (T$_{in}$) to vary and assume a steady state/constant 
mass-loss rate density distribution for all of the stars. Their adopted temperatures are similar to ours.  Despite the difference in the DUSTY 
modeling, and our adopted distance is also somewhat less than they used, the luminosities for   the four less-luminous RSGs  
 agree within our quoted errors although our $\dot{M}$ are somewhat lower. The low errors on the luminosities cited by  \citet{BD2018} did not include the distance uncertainty \citep{DB2018}.  We derive a higher luminosity and mass-loss rate for MY~Cep  
probably due to the  additional contribution from the circumstellar 
dust with the non-constant mass loss rate.

\citet{BD2016,BD2018} argue that a variable dust condensation 
temperature (T$_{in}$) gives better fits to the SEDs, and suggest that stars with lower $\dot{M}$ have a larger spread in T$_{in}$.  The dust formation temperature however depends only on the dust chemistry, vapor pressure, and properties of the grains \citep{Gail} and is independent of the mass-loss rate. The temperature of the inner shell, T$_{in}$, however, via r$_{min}$ will alter the $\dot{M}$. Given the lack of a physical mechanism for similar stars to have significantly different dust condensation temperatures, we adopt the same T$_{in}$  at 1000K for all the stars.  A lower T$_{in}$, such as 750K, within the expected range for silicate dust, will approximately double the mass loss rate.

\subsection{RSGC3} 

RSGC3, one of the three highly obscured clusters discovered near the base of the Scutum/Crux spiral feature has, on average, the lowest luminosity population of red supergiants among
the four clusters discussed in this paper. Only four of its probable members
shows evidence for significant circumstellar dust at the long wavelengths. Two independent
investigations by \citet{Alexander} and by \citet{Clark} announced the discovery
of this cluster at about the same time. In this paper we use the star numbers from \citet{Clark} because they give a longer list of members and candidates. 

\citet{Clark} list 15 members in their core group based on the photometry plus 7 additional likely members. They also include a third group described as ``stars of interest'' based on their magnitudes and colors. \citet{Alexander} list 8 probable members, 6 candidates and 2 foreground stars. Their 8 members are in the \citet{Clark} core group. Only one star, (RSGC3-10) included in the core group by \citet{Clark}, is considered a candidate by \citet{Alexander}.  
In addition to the published 2MASS and Glimpse photometry, both groups also obtained low to moderate resolution near-infrared spectroscopy for spectral classification and for confirmation of the  stars' red supergiant status based on the CO band head absorption.

We measured long wavelength fluxes for 9 members of the core group that were visible in the SOFIA/FORCAST frame.  The fluxes for these stars in the Appendix range from 5.5 to 11.1$\mu$m. They were not detected at longer wavelengths even though those images were observed.  For our analysis with DUSTY, we added 11 stars that are considered members and 
candidates with mid to far-IR fluxes from other sources.

The greatest uncertainty in our analysis and subsequent discussion of the stars in RSGC3  is the adopted distance and therefore the luminosity of the stars, and their mass-loss rates.  \citet{Clark} suggest a distance of 6 $\pm$ 1 kpc based on the cluster's location at the tip of the Galactic bar, similar to RSGC1 and RSGC2, with a possible maximum distance of 7.2 kpc based on the measured interstellar extinction. \citet{Alexander} also suggest a distance of 7.2 kpc based on the adopted mean colors and corresponding absolute magnitudes. 

As already emphasized above, our adopted distance of 6 $\pm$ 1 kpc distance leads to an uncertainty of $\approx$ 37$\%$  in the derived luminosities. The spectral types in these two papers differ by an average of $\pm$  one and a half  spectral sub-types, or approximately $\pm$ 100~K corresponding to an uncertainty of 10\% in the luminosity.  Both \citet{Clark} and \citet{Alexander} estimate a mean A$_{K}$ of 1.5 mag corresponding to A$_{V}$ of 12.6 to 13.0 mag. Although the two extinction estimates agree, we assume an error of  $\pm$ 0.1 mag in A$_{K}$,or 7\% on the luminosity, and the residuals of the fit to the SEDs, average 6\%.  Combining the errors gives an uncertainty of 43\% in the luminosity and 28\% in the mass loss rate.  

Our derived parameters and the results of the DUSTY modeling are summarized in 
Table 6 for 20 probable and candidate members. Comparison of our luminosities with those for eight stars in common with \citet{Clark} agree within the quoted errors in both papers, but we note that our luminosities are systematically lower by 0.12 dex, on average.

\begin{deluxetable*}{llccclll}[htbp]  
\tabletypesize{\footnotesize}
\tablenum{6}
\tablecaption{Model Results for RSGC3 Red Supergiants\tablenotemark{a}}
\tablewidth{0pt}
\tablehead{
\colhead{Star} & 
\colhead{Sp Type/Temp} & 
\colhead{Power law(r $^{n}$)} & 
\colhead{$\tau_{v}$} &
\colhead{r$_{1}$ (AU)} & 
\colhead{L$_{\odot}$} &
\colhead{M$_{\odot}$ yr$^{-1}$} & 
	\colhead{Comment\tablenotemark{b}} 
}
\startdata 
RSGC3-1*  & \nodata/3600 & 2.0 & 0.1 & 27 & 3.7 $\pm$ 1.6 $\times$ 10$^{4}$ & 1.7 $\pm$ 0.5 $\times$ 10$^{-7}$& A-4, M0 \\
RSGC3-2*  & M3 Ia/3800 & 2.0 & 0.07 & 25 & 2.9 $\pm$ 1.0 $\times$ 10$^{4}$ & 1.1  $\pm$ 0.4 $\times$ 10$^{-7}$ & A-8, K5\\
	RSGC3-3   & M4 Ia/3500  & 1.5  & 0.1 & 37  & 7.3 $\pm$ 3.1 $\times$ 10$^{4}$ & 2.4 $\pm$ 0.7 $\times$ 10$^{-6}$  &  A-6 M4, (2.3 $\times$ 10$^{-7}$) \\
	RSGC3-4*  & M3 Ia/3600 & 1.7 & 0.1 & 32 & 5.0 $\pm$ 2.1 $\times$ 10$^{4}$ & 8.2 $\pm$ 2.3 $\times$ 10$^{-7}$ & A-7, M2, (1.9 $\times$ 10$^{-7}$) \\
RSGC3-5*  & M2 Ia/3800 & 1.7 & 0.1 & 31 & 4.7 $\pm$ 2.0 $\times$ 10$^{4}$ & 8.1 $\pm$ 0.5 $\times$ 10$^{-7}$ & A-3, M1\\
RSGC3-6*  & RSG/4000   & \nodata  & \nodata & \nodata &  2.7 $\pm$ 1.1  $\times$ 10$^{4}$ & \nodata &  A-1, K5, no IR excess  \\
RSGC3-7*  & M0 Ia/3800 & 2.0 & 0.2 & 26 & 2.9 $\pm$ 1.2 $\times$ 10$^{4}$ & 3.1  $\pm$ 0.9 $\times$ 10$^{-7}$ & A-5, K4 \\
RSGC3-8   & K5 Ia/3800 &\nodata  & \nodata  &\nodata  & 1.8  $\pm$ 0.8 $\times$ 10$^{4}$ & \nodata  & A-12, cand., no IR excess \\
RSGC3-9   & M0 Ia/3800 & \nodata  & \nodata  &\nodata  & 2.8 $\pm$ 1.2 $\times$ 10$^{4}$ & \nodata  &  A-9, K4, no IR excess \\
RSGC3-10*  & M0 Ia/3800 & \nodata  & \nodata  &\nodata  & 1.0 $\pm$ 0.4 $\times$ 10$^{4}$ & \nodata & A-10, cand., no IR excess \\
RSGC3-11   & RSG/4000  & \nodata  & \nodata  &\nodata  & 2.7 $\pm$ 1.1 $\times$ 10$^{4}$ & \nodata & no IR excess \\ 
RSGC3-12   & \nodata/4000 & 2.0  & 0.1 & 28 & 1.7 $\pm$ 0.7 $\times$ 10$^{4}$ & 1.1 $\pm$ 0.3 $\times$ 10$^{-7}$ & \nodata    \\ 
	RSGC3-13   & RSG/3600   & 1.7  & 0.1 & 36  & 6.5 $\pm$ 2.8 $\times$ 10$^{4}$  &  9.4 $\pm$ 2.6 $\times$ 10$^{-7}$ & \nodata, (2.2 $\times$ 10$^{-7}$)  \\ 
RSGC3-14    & RSG/3300   & 2.0  & 0.1  & 22  & 2.4 $\pm$ 1.0 $\times$ 10$^{4}$ &  1.3 $\pm$ 0.4 $\times$ 10$^{-7}$  & A-16, cand. \\
	RSGC3-15    & RSG/3200 & 1.3 & 0.4 & 29  & 5.0 $\pm$ 2.0 $\times$ 10$^{4}$ &  7.9 $\pm$ 2.2 $\times$ 10$^{-6}$ &    \\
RSGC3-16    & RSG/3200 & 2.0 & 0.1 & 30 & 4.4 $\pm$ 1.9 $\times$ 10$^{4}$ & 1.8 $\pm$ 0.5 $\times$ 10$^{-7}$ &   \\
RSGC3-17    & \nodata  & 1.5 & 0.7 & \nodata &\nodata & \nodata &  see text \\
RSGC3-21     & \nodata  & 1.5  & 0.1  & \nodata  & \nodata & \nodata  &  see text\\
RSGC3-27*     & \nodata  & 1.7 & 0.1 &\nodata  &\nodata  & \nodata  &  A-13, cand. see text \\
RSGC3-A-11*    & \nodata  & 1.6  & 0.1 & \nodata & \nodata  & \nodata &  cand. see text\\ 
\enddata
\tablenotetext{a}{Stars with  FORCAST fluxes have an asterisk.}
	\tablenotetext{b}{The Comment includes the designation and spectral type from \citet{Alexander} and $\dot{M}$ for a constant mass loss rate, n $=$ 2, for those cases where the power law discrimination  is less certain.}
\end{deluxetable*}

The most luminous members show evidence in their SEDs for circumstellar dust with silicate emission features, not apparent in the other member stars.  RSGC3-15, however, is exceptional with a very strong silicate emission and higher optical depth. Although a spectral type is not available, its relative low temperature of 3200~K is consistent with significant circumstellar dust. Its SED is shown in Figure 6 together with  RSGC3-13. The latter's SED illustrates the uncertainty in adopting the best power law fit and the $\dot{M}$ especially with the lack of longer wavelength fluxes. For this star and two others, we include the $\dot{M}$ for constant mass loss in the comment column in Table 6. Five of  the  members have little or no circumstellar dust like the less luminous RSGs in N7419. For these stars no $\dot{M}$ is given in the Table.

\begin{figure}[h]  
\plottwo{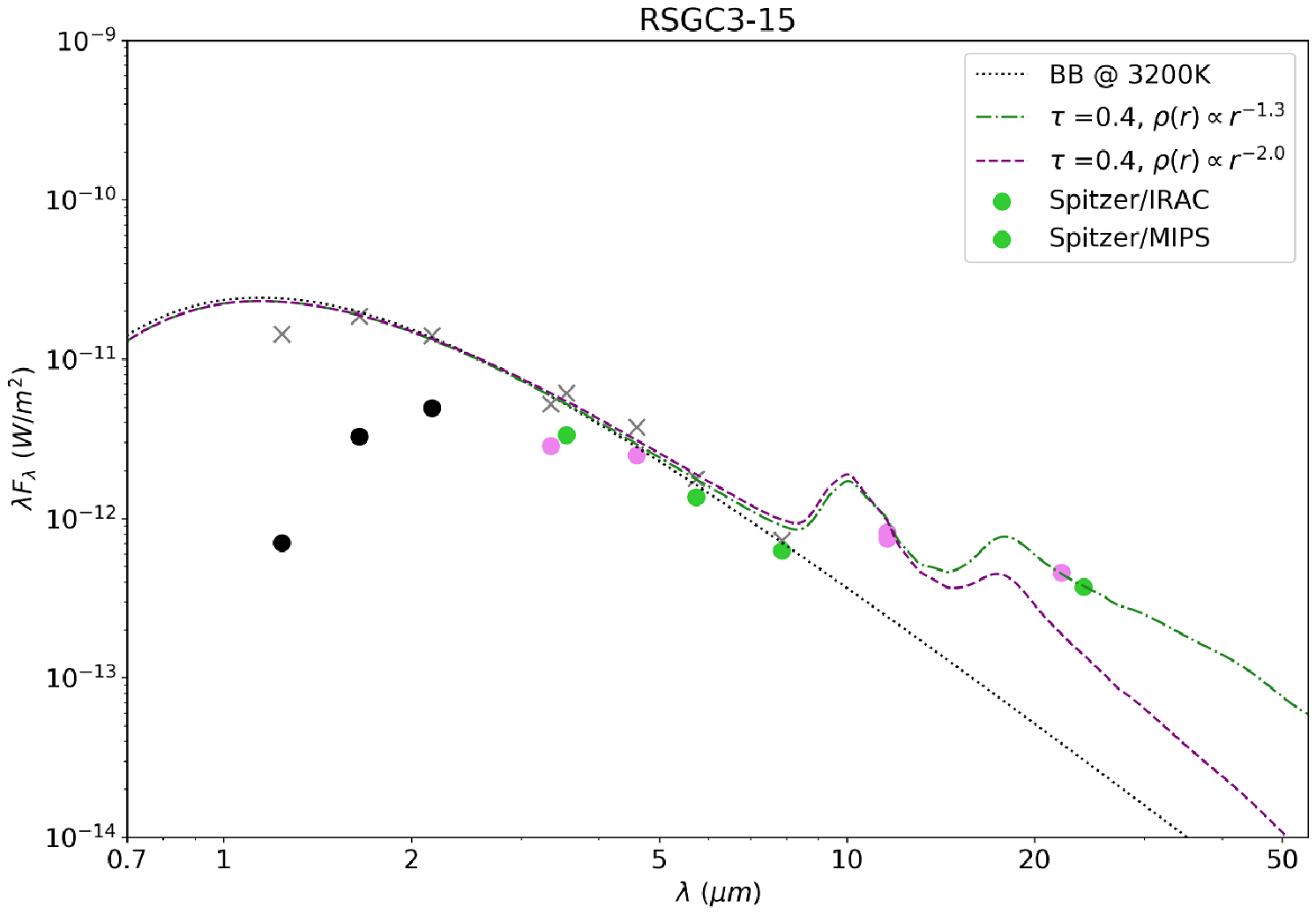}{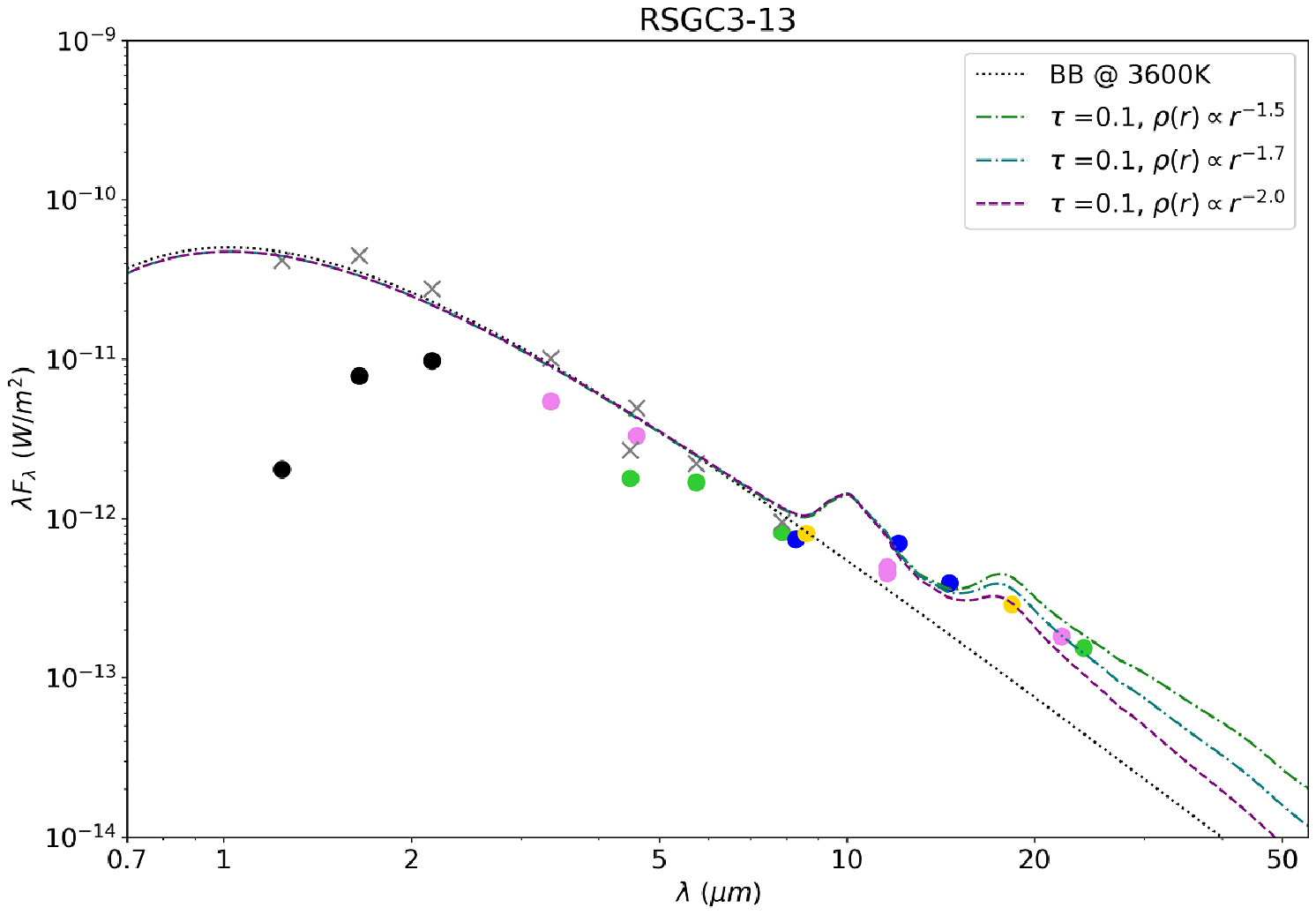}
\caption{Left: The observed and extinction corrected fluxes for RSGC3-15, one of the most luminous members, with very  strong silicate emission due to circumstellar dust. The  DUSTY fits support a higher $\tau$, and a less steep power.  Right: The SED for RSGC3-13 is an example of the uncertainty in the power law fits with the DUSTY models. The color-coded symbols are the same as in Figure 5 with the addition of Spitzer/IRAC and MIPS fluxes.}  
\end{figure}

Four of the stars included in Table 6, however are anomalous  
in the sense that they are much redder than the other members which may be due to higher interstellar extinction, some possible circumstellar reddening, or 
they may not be members, but background stars. RSGC3-27 (A13) and A11 are both 
located in the core or central region of the cluster (see Figure 1 in \citealt{Alexander}) and were thus on our SOFIA/FORCAST frame. Both are noteworthy as the reddest stars in the two-color diagram in \citet{Alexander} who consider them candidate members. 
\citet{Clark} does not include A11, but considers RSGC3-27 to be a ``star of interest''. Their SEDs, corrected for the adopted interstellar extinction, imply temperatures below 3000~K. Thus they likely have higher extinction. Assuming that they are RSGs, and estimating their extinction from the observed colors, the resulting SEDs imply unrealistic high temperatures.
We note however that the reddening free parameter  Q$_{IR}$\footnote{Q$_{IR}$ $=$ (J-H) $- 1.8 \times$ (H-K$_{s}$) \citep{Clark}} for RSGC3-27 of 0.1, suggests that it may indeed be a an earlier-type star; the Q$_{IR}$ for A11 of 0.35 is consistent with an RSG. Neither star has a large 10--$20\,\mu$m silicate emission feature 
which probably eliminates circumstellar dust as a contributor to the colors. 

RSGC3-17 and 21 are similarly very red stars, and are considered likely members by 
\citet{Clark}. Like RSGC3-27 and A11,  their colors and SEDs suggest that they have higher interstellar extinction. Their  Q$_{IR}$ parameters support their 
classifiction as RSGs, but 
their SEDs corrected for the adopted extinction for RSGC3 imply temperatures even below 2000~K. Given their positions away from the center of the cluster, we suggest that they 
may both be background stars. RSGC3-17 has a significant silicate emission 
feature, 
the largest observed for the RSGC3 stars, and possible additional excess 
radiation in the 3--$8\,\mu$m region. Therefore it very likely has some 
additional circumstellar extinction unlike the other stars in RSGC3, and may not be a member. 

The SEDs for RSGC3-17 and A11 are shown in Figure 7.  Spectra and radial velocities are needed for these four stars to confirm or not their membership in RSGC3.
Due to these uncertainties, they are not included in later discussion.

\begin{figure}[h]   
\plottwo{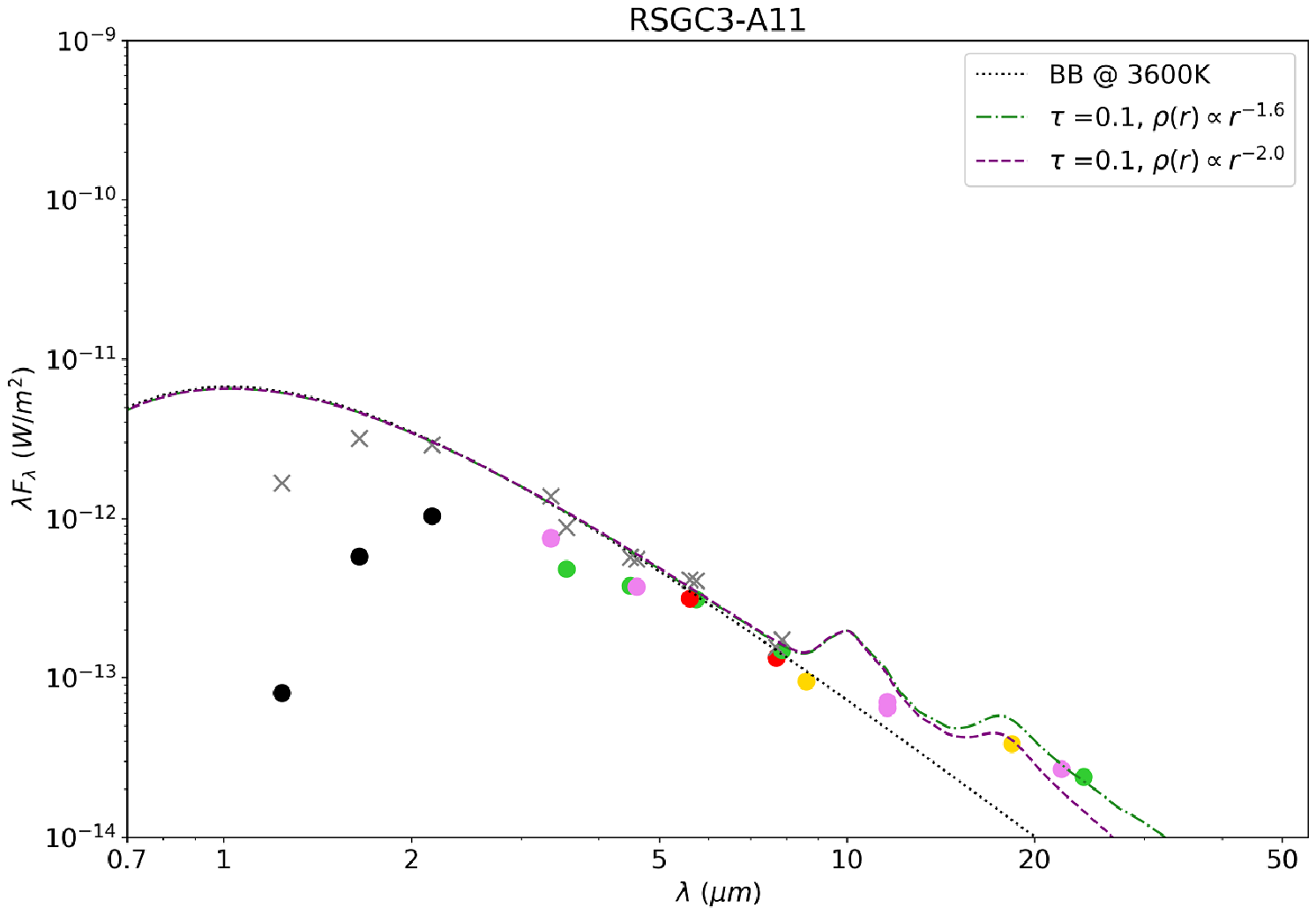}{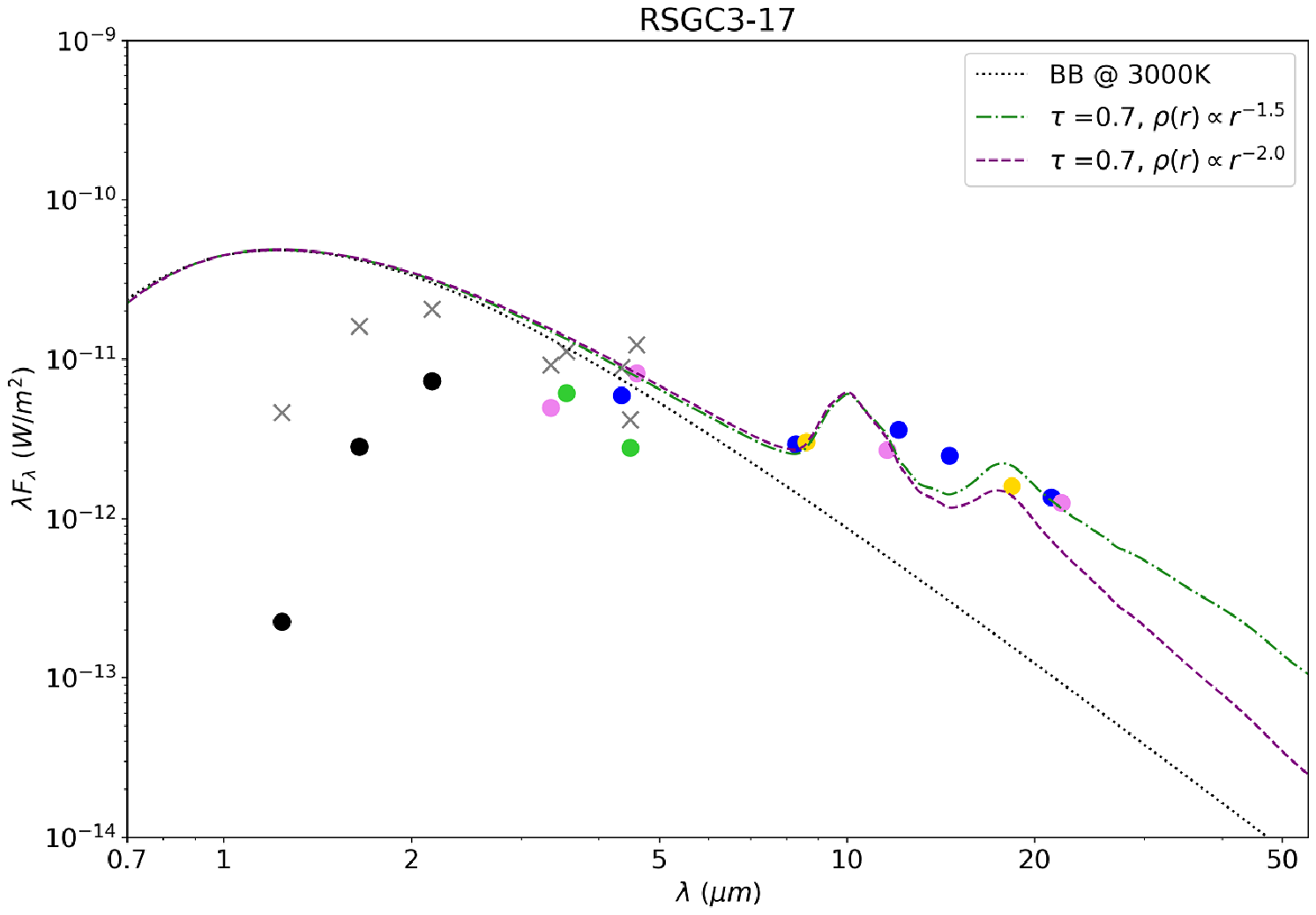}
	\caption{The SEDS for RSGC3-A11 (left) and  RSGC3-17 (right).   
	Both stars are very red suggesting that they have have additional extinction. Planck curves  with temperatures consistent with their spectral types do not fit the observed SEDs corrected for the adopted interstellar extinction. A11 would require a temperature below 3000~K and RSGC3-17 below even 2000~K. It has a significant silicate emission feature and possible excess radiation in the mid-infrared quite different from the other RSGC3 red supergiants and may be a 
	background star. See text for more discussion. The symbols are the same as in Figures 5 and 6.}
\end{figure}

\subsection{RSGC2}

RSGC2 is the second highly reddened cluster with an unprecedented number of
red supergiant members and candidates \citep{Davies07}, and like the other two 
it is located at the base of the Scutum-Crux spiral arm. It was first noticed by \citet{Stephenson} who commented on the number of red stars. Hence it is also known as Stephenson 2. 
\citet{Davies07} obtained near-infrared spectra for classification  based on the CO bands and radial velocities and compiled the available infrared photometry from 2MASS, Glimpse, and MSX for discussion of their luminosities and the foreground extinction. Based on the spectral types and velocities they identify a core group of 26 probable RSG members. \citet{Negueruela12,Negueruela13} reported far-red 
spectra for a subset of the stars with improved spectral types based on atomic-line  and molecular-band strengths.

\citet{Davies07}  derive a kinematic distance of 5.8 kpc $\pm \approx$ 1 kpc and estimate a mean foreground extinction of A$_{K} =$ 1.44 mag (A$_{V} =$ 12.9 mag) from the observed colors of the individual stars. As the authors note, several of the candidate members have much higher extinction and in some cases it may include a contribution from circumstellar dust. We therefore re-measure the mean A$_{K}$ deleting several stars, derive a mean of 1.32 mag $\pm$ 0.13 mag from 18 stars (A$_{V} =$ 11.9 mag), and adopt a 10\% error on the luminosity from the uncertainty in the foreground extinction.  The combined error on the luminosity and mass loss rate for RSGC2 is the same as for RSGC3.  

RSGC2 is extended over about 6$\arcmin$  and would require two pointings with FORCAST to include the majority of the RSG members. We  made a single pointing centered on the highest-density concentration of RSGs around stars 14 and 15. Our FORCAST measurements for the 10 RSGs in the frame are in Table B3.  The long wavelength fluxes range from 7.7 to 37.1$\mu$m. In addition to these 
10 stars we added 5 RSG members with mid- to far-infrared fluxes from the other sources for inclusion in our DUSTY modeling and analysis.  Our derived parameters and results are summarized in Table 7 for 15 red supergiants. Our luminosities agree quite well for eight stars in common with \citet{Davies07}; well within
the quoted errors and with no systematic difference.

\begin{deluxetable*}{llccclll}[htbp]   
\tabletypesize{\footnotesize}
\tablenum{7}
\tablecaption{Model Results for RSGC2 Red Supergiants\tablenotemark{a}}
\tablewidth{0pt}
\tablehead{
\colhead{Star} & 
\colhead{Sp Type/Temp} & 
\colhead{Power law(r $^{n}$)} & 
\colhead{$\tau_{v}$} &
\colhead{r$_{1}$ (AU)} & 
\colhead{L$_{\odot}$} &
\colhead{M$_{\odot}$ yr$^{-1}$} & 
\colhead{Comment} 
}
\startdata 
	RSGC2-2*  & M3(M7/3200)\tablenotemark{b} & 1.5 & 0.4 & 52 & 1.6 $\pm$ 0.7 $\times$ 10$^{5}$ & 1.3 $\pm$ 0.4 $\times$ 10$^{-5}$  & Stephenson 2 \\
	RSGC2-3*  & M4(M5/3400) & 2.0 & 0.2 & 46 & 8.8 $\pm$ 3.8 $\times$ 10$^{4}$ &  5.1 $\pm$ 1.4 $\times$ 10$^{-7}$ & Stephenson 10\\
	RSGC2-5   & M4(M5/3400)  & 1.5  & 0.3 & 47 & 1.0 $\pm$ 0.4 $\times$ 10$^{5}$ & 5.8 $\pm$ 1.6 $\times$ 10$^{-6}$  &  \nodata \\
	RSGC2-6* & M5(M3.5/3600) & 2.0 & 0.1 & 31 & 5.3 $\pm$ 2.3 $\times$ 10$^{4}$ & 2.0 $\pm$ 0.6 $\times$ 10$^{-7}$ & Stephenson 1\\
RSGC2-8*  & K5/3900 & 2.0 & 0.1 & 43 & 8.4 $\pm$ 3.4 $\times$ 10$^{4}$ & 2.6 $\pm$ 0.7 $\times$ 10$^{-7}$ & Stephenson 4\\
RSGC2-10*  & M5/3500   & 2.0 & 0.1 & 37 &  7.2 $\pm$ 3.1 $\times$ 10$^{4}$ & 2.2 $\pm$ 0.6 $\times$ 10$^{-7}$ & \nodata  \\
RSGC2-11*  & M4/3600 & 2.0 & 0.1 & 31 & 4.9 $\pm$ 2.1 $\times$ 10$^{4}$ & 1.9 $\pm$ 0.5 $\times$ 10$^{-7}$ & \nodata \\
RSGC2-13   & M4/3700 &2.0  & 0.08 & 30  & 4.2 $\pm$ 1.8 $\times$ 10$^{4}$ & 1.4 $\pm$ 0.5 $\times$ 10$^{-7}$ & \nodata \\
RSGC2-14*   & M3/3600 &2.0  & 0.1  & 23 & 2.7 $\pm$ 1.2 $\times$ 10$^{4}$ & 1.4 $\pm$ 0.4 $\times$ 10$^{-7}$ & Stephenson 5\\
RSGC2-15*  & M2/3700 & \nodata  & \nodata  &\nodata  & 1.4 $\pm$ 0.6 $\times$ 10$^{4}$ & \nodata & Stephenson 6, no IR excess \\
RSGC2-17   & K3/4000  & \nodata  & \nodata  &\nodata  & 4.7 $\pm$ 2.0 $\times$ 10$^{4}$ & \nodata & no IR excess   \\ 
	RSGC2-18*   & M4(M0.5/3800) & \nodata  & \nodata  &\nodata  & 5.3 $\pm$ 2.3 $\times$ 10$^{4}$ & \nodata & Stephenson 7, no IR excess\\ 
	RSGC2-23   & M4/3500   & 2.0  & 0.1 & 34  & 5.9 $\pm$ 2.5  $\times$ 10$^{4}$ & 2.1 $\pm$ 0.9 $\times$ 10$^{-7}$  & pec. SED \tablenotemark{c}, non-member?  \\ 
	RSGC2-49   & K4/4000  & 1.3 & 5 & 115  & 3.9 $\pm$ 1.7 $\times$ 10$^{5}$ & 7.7  $\times$ 10$^{-4}$   & large IR  excess, see text \\
\nodata  & \nodata   & 2.0  & 14.5  & 145  & \nodata & 1.3 $\times$ 10$^{-4}$   &    \\
	RSGC2-52*    & M0/3800 & 1.5 & 0.1  & 23  & 2.4 $\pm$ 1.0 $\times$ 10$^{4}$  & 1.4 $\pm$ 0.6 $\times$ 10$^{-6}$ &  pec SED \tablenotemark{c}  \\
\enddata
\tablenotetext{a}{Stars with  FORCAST fluxes have an asterisk.}
	\tablenotetext{b}{The improved spectral types from \citet{Negueruela12,Negueruela13} are given in parenthesis. These types and temperatures are used in our discussion and analysis}
	\tablenotetext{c}{Although our Planck curve fits to the SEDs for RSGC3-12 and 52 yielded resonable results, they failed to fit the J-band fluxes for the  expected range of temperatures. \citet{Davies07} derive somewhat higher A$_{K}$ values for these two stars which may be due to circumstellar extinction or variable extinction in the field. Using these higher values, however, as in the case of examples in RSGC3 like A11, gave  temperatures inconsistent with the spectral types. We suggest that based on its radial velocity and position, RSGC2-23 may be a background star with higher extinction, but RSGC2-52 is in the core of the cluster. \citet{Negueruela12} also suggests RSGC2-23 is a non-member.}  
\end{deluxetable*}

RSGC2 has a more luminous population of red supergiants and although the range of their luminosities overlap those in RSGC3, this cluster includes at least three  stars with luminosities above  10$^{5}$ L$_{\odot}$, and with corresponding higher mass loss rates  including a very late and rare M7, RSGC2-02 and a probable post-red supergiant, RSGC2-49. The SEDs for the majority of the stars also reveal  silicate emission at the long wavelengths due to circumstellar dust. The SEDs for two of the most luminous stars are shown in Figure 8.  

\begin{figure}[h]   
\epsscale{1.0}
\plottwo{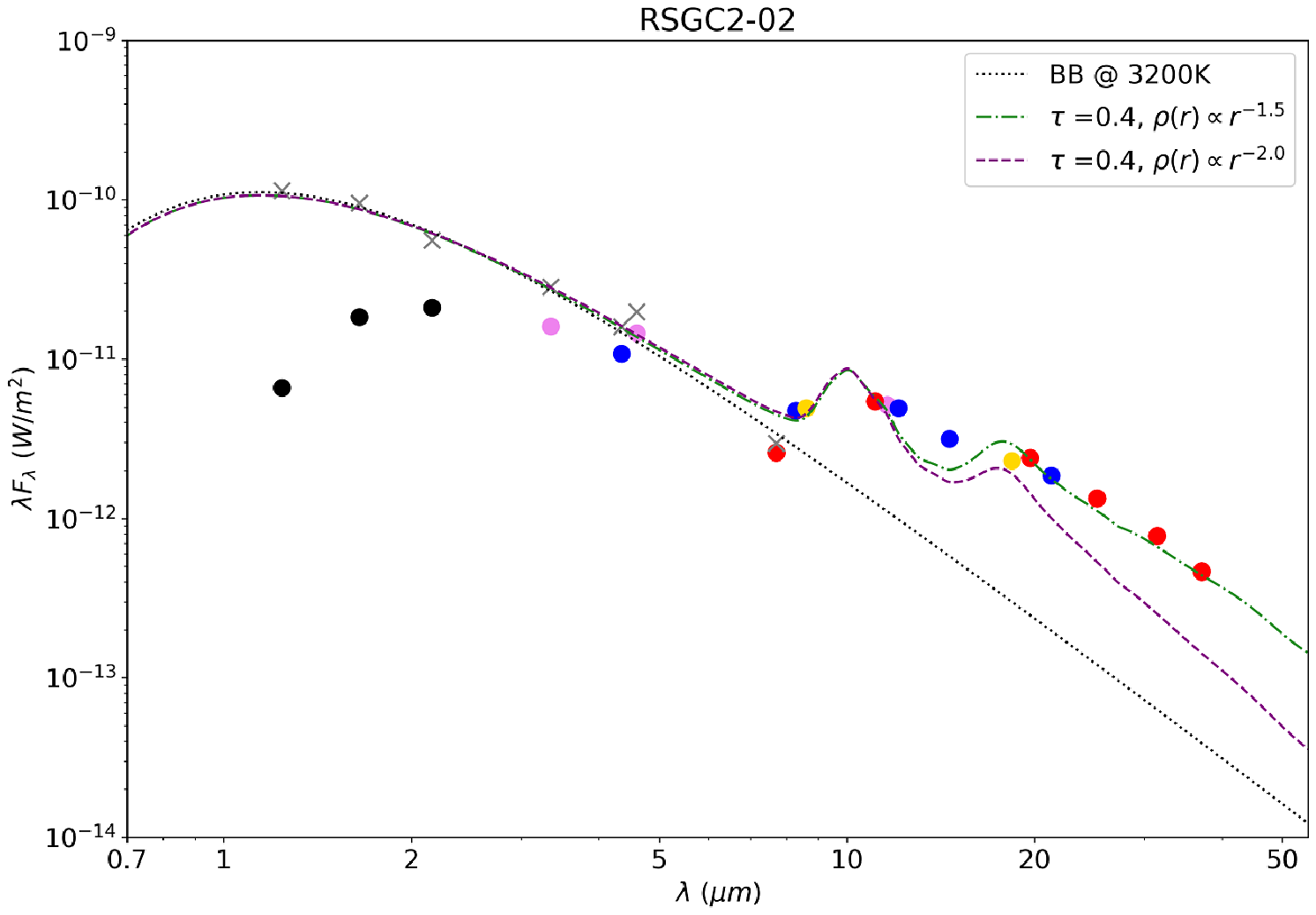}{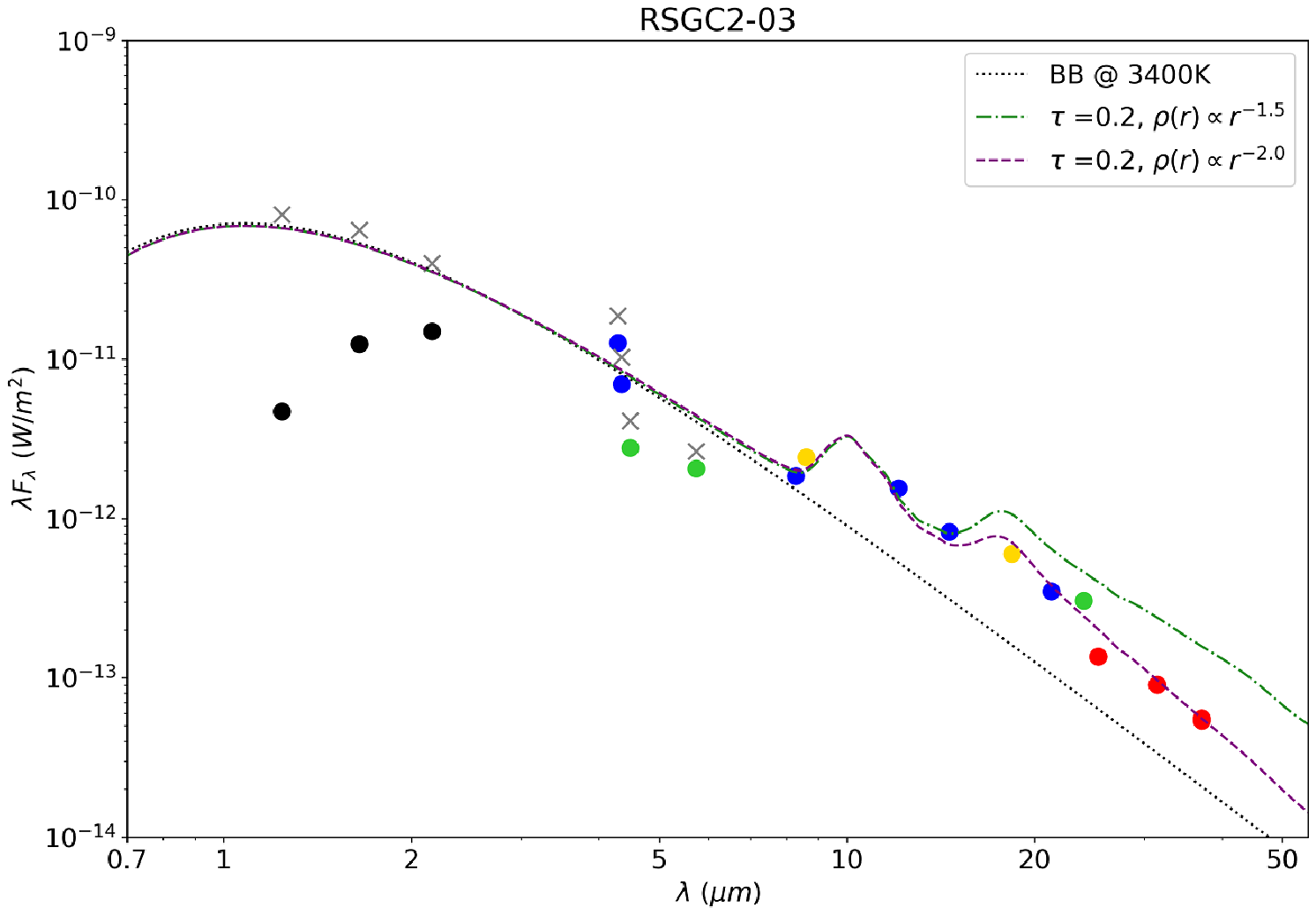}
	\caption{RSGC2-02 (left) and RSGC2-03 (right)  are two of the most luminous red supergiants in RSGC2, but  long wavelngth fits from DUSTY  yield somewhat different models for their mass loss. RSGC2-02 is the M7 red supergiant which very likely accounts for its strong silicate emission and high mass loss rate. RSGC2-3 is also a late-type M5 and has strong silicate emission but shows a good 
	fit with a constant mass loss rate. These two stars have obviously had different mass loss histories.  The symbols are the same as in Figures 5 and 6.} 
\end{figure}

There are  a surprisingly  large of number of very late-type, M4, M5 and an M7, supergiants in RSGC2. This was also noted by \citet{Davies07}. Red supergiants with these late spectral types are relatively rare in the Solar neighborhood \citep{RMH,Elias}. The prominence of earlier M spectral types in the Magellanic Clouds, compared to our Solar neighborhood, for example,  is attributed to the reduced metallicity and therefore lower opacity in the atmospheres of the RSGs in 
the Clouds. So perhaps the shift to later types
in RSGC2 is due the expected higher metallicity towards the Galactic center, although the same shift to later M types is not apparent in RSGC3. Red supergiants with these later M types are also associated with enhanced circumstellar dust and high mass loss and thus may be related to the evolutionary state of the stars. 

\citet{Negueruela13} have suggested that RSGC1-01 may be a member. Their spectra show a very late-type RSG similar to RSGC2-02. The star's radial velocity is slightly offset from the other members by about 20 km s$^{-1}$. It is located away from the central region of the cluster and it is by far the brightest star in the field in the K-band image (see Figures 1 and 4 in \citet{Davies07}).  It is not in the FORCAST FOV, so we determined its SED from published fluxes. Adopting the distance to RSGC2, the mean extinction, and integrating is SED yields 6.3 $\times 10^{5} L_{\odot}$ which would make it clearly the most luminous member of the cluster, and much more luminous than RSGC2-02 and the post-RSG star RSGC2-49. But its SED is somewhat peculiar. With the adopted extinction, the 2MASS J and H band fluxes cannot be fit with a range of appropriate temperatures for an RSG.  Their fluxes suggest a higher extinction. If so RSGC1-01 would be even more luminous.  Although these  properties do not rule out  membership in the cluster, we consider it doubtful. As emphasized by \citet{Negueruela12}, the region of the RSG clusters is complex, extended spatially across and along the line of sight.

The cluster's most interesting star is RSGC2-49. Its observed SED (Figure 9) clearly shows the evidence for high circumstellar extinction and high mass loss with significant infrared excess radiation from 8 to $> 20\,\mu$m. The lack of a silicate emission feature in the 10--$20\,\mu$m region indicates that the emission is optically thick. Based on its apparent early K spectral type, luminosity and position in the HR Diagram (see \S {5.3}), RSGC2-49 is a post-red supergiant similar to IRC~+10420 and IRAS 17163-3907 in the Galaxy \citep{RMH97,Tiffany,Koumpia} and Var~A in M33 \citep{RMH06}.

Star 49's SED closely resembles that of the famous red supergiant VY CMa \citep{Shenoy16} although it is much warmer star. \citet{Davies07} derived a very high A$_{K}$ of 4.6~mag due to a combination of interstellar and circumstellar extinction.    Ideally we would correct its SED for the interstellar extinction component and integrate to the longest wavelengths to
account for the flux re-radiated by the dust and responsible for the additional 
circumstellar extinction. Unfortunately longer wavelength fluxes from FORCAST or other sources to better model is energy distribution are not available to estimate its total luminosity, unlike for VY~CMa. Instead, we integrate a 4000~K Planck curve fit through the K-band flux corrected for the A$_{K}$ of $\approx$ 4.6~mag which should  approximate its SED when corrected for the flux reradiated by the dust at longer wavelengths. The total luminosity estimated in this way is 3.9 $\times 10^{5} L_{\odot}$. This may be an under-estimate. It does not include possible excess radiation from warm dust in the 2--$8\,\mu$m 
region. This luminosity is slightly higher than that from \citet{Davies07} based on the absolute K-band magnitude. 

Figure 9 shows the results for two DUSTY models relative to its optically-thick silicate emission. Both require high values of $\tau_{v}$, and both yield  high 
values of the mass loss rate, on the order of 1.3 to 7 $\times$ 10$^{-4}$ M$_{\odot}$ yr$^{-1}$.  These are not exceptional mass loss rates for hypergiants. Similar rates are  measured for stars like IRC~+10420, Var A and for red supergiants like VY CMa and NML Cyg.  As a probable post-RSG star, we suspect that star 49 is experiencing constant as well as possible variable mass loss in the past as observed for 
IRC~+10420 \citep{Shenoy16}.

\begin{figure}[h]   
\epsscale{0.7}
\plotone{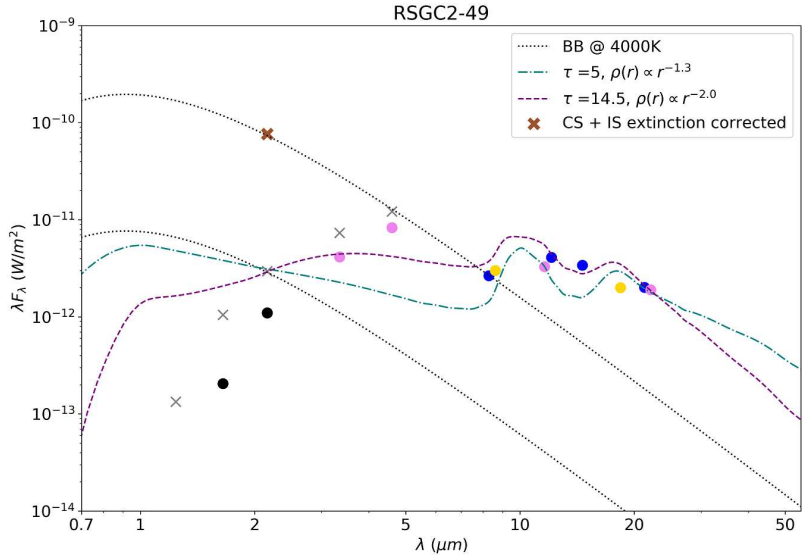}
	\caption{The SED for RSGC2-49. Its SED clearly reveals a large infrared excess and the impact of high circumstellar extinction on the observed fluxes.  We show the observed fluxes corrected for the mean interstellar extinction adopted for RSGC2 with a 4000K Planck curve fit through the K-band flux. The red tipped cross shows the K band flux corrected for the A$_{K}$ extinction from \citet{Davies07} attributed to both interstellar and circumstellar dust. The silicate emission feature is optically thick; the  figure shows the results from two DUSTY models  fit through the observed K-band flux corrected for foreground interstellar extinction. See the text for more discussion. The symbols are the same as in the previous SEDs.}   
\end{figure}

\subsection{RSGC1} 

RSGC1 is the youngest of the clusters and its RSG members are the most luminous.  Near-infrared spectra in the discovery paper by \citet{Figer} and in the 
follow-up paper \citep{Davies08} provide confirmation of the red supergiant status of 15 stars based on the strength of the CO band heads. All of the members (except star 14) discussed here have luminosities above 10$^{5}$ L$_{\odot}$ and with, on average, higher mass loss rates than measured for the red supergiants in the other clusters. RSGC1 is thus critical for establishing the shape of the mass loss rate-luminosity relation for the most luminous RSGs. 

Like RSGC2 and RSGC3, RSGC1 is located near the junction of the Scutum-Crux spiral arm and the tip of the Galactic bar.  
Published distances for RSGC1 range from 5.8 kpc \citep{Figer} to 6.6 kpc \citep{Davies08} $\pm$ 0.9 kpc. It may be the most distant of the four clusters and also suffers from very high and uncertain interstellar extinction. The two references above, respectively estimate A$_{K}$  values of 2.74 and 2.60 mag which imply visual extinction as high as 25 to 30 mag. Almost all of these stars show evidence in their SEDs for circumstellar dust and extinction in addition to the interstellar component. To separate the interstellar component, \citet{BD2020} select RSGC1-14, the lowest luminosity RSG in the cluster, assume no additional circumstellar extinction, and determine A$_{V}$ of 25 $\pm$ 2 mag. We adopt  A$_{K}$ of 2.29 $\pm$ 0.2~mag for this same star for the mean interstellar extinction in this paper. Although we suspect that the extinction correction is more uncertain, it contributes a 10\% uncertainty  to the luminosity as in the other clusters. The difference in the published spectral types from \citet{Figer} and from \citet{Davies08} is $\pm$ two subtypes, or $\approx$ $\pm$200~K. There is more scatter in the fluxes from the different sources for the stars in this cluster which may be 
due to the faintness of the stars or to variability in these more luminous RSGs. As a result, highly discrepant points were deleted from the fits to the SED and the error from the residuals is larger than in the other clusters and more variable from star to star. An uncertainty of $\approx$ 15\% on the derived luminosity is typical. The combined error is 47\% in luminosity and 28\% in the mass loss rate.  

The SEDs of all of the stars observed with FORCAST show significant silicate 
emission features and extended emission longward of 20~$\mu$m (except the yellow hypergiant RSGC1-15). Figure 10 shows the SEDs for two of the members. The results of our DUSTY modeling are summarized in Table 8.The spectral types from the strength of the CO bands  from \citet{Figer} and from \citet{Davies08} are both listed, in that order, with our adopted temperature for the SED fits.  

\begin{figure}[h]   
\plottwo{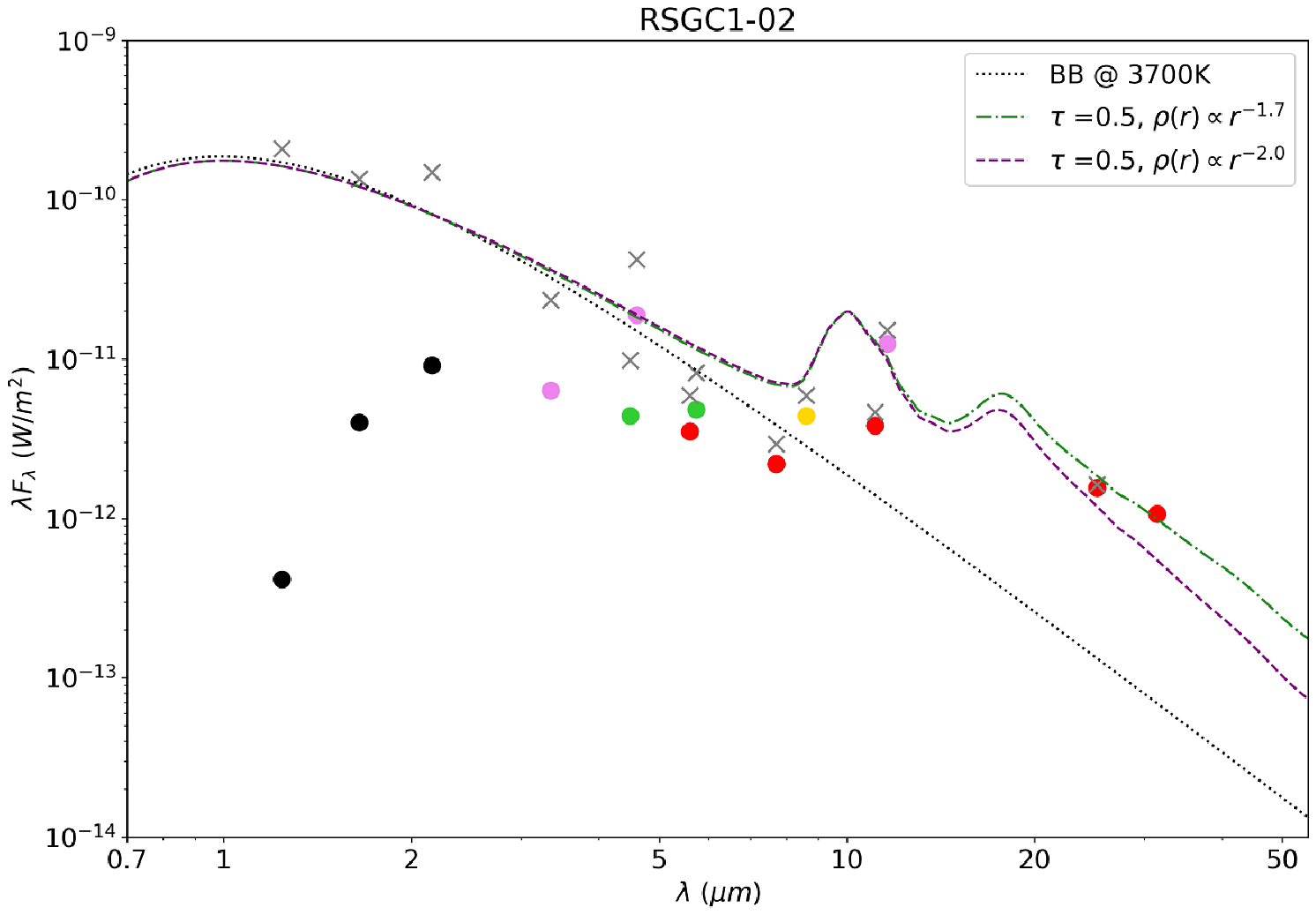}{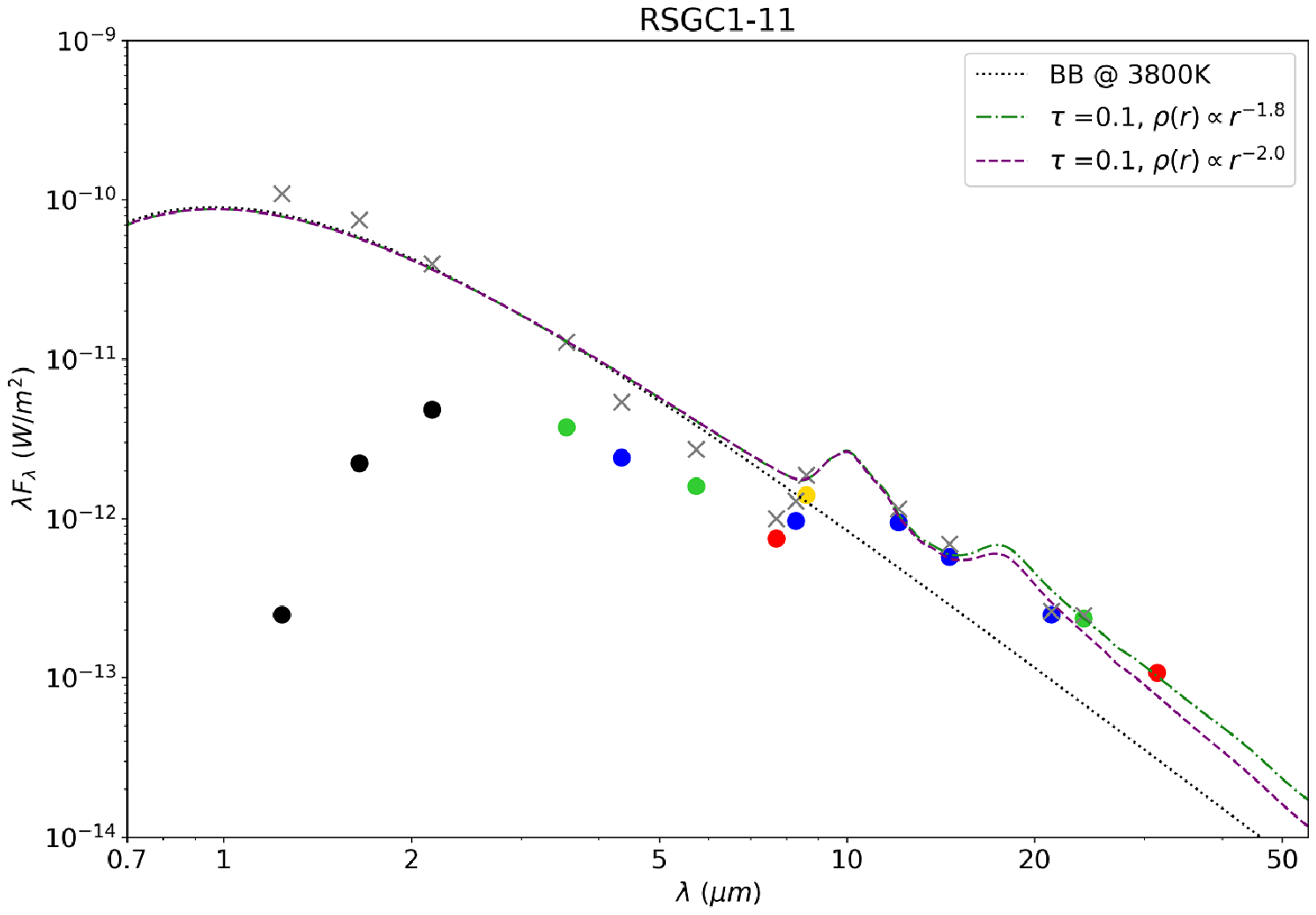}
\caption{Left:RSGC1-02 is one on the most luminous members with SiO maser emission and a prominent silicate emission feature with a corresponding high mass loss rate. Right: RSGC1-11 has a comparable luminosity and while it also has  silicate emission, it is weaker with a lower mass loss rate. The SEDs for both stars illustrate the scatter in the observed fluxes from different sources for the stars in this cluster. In our adopted fits from DUSTY we emphasize the longest wavelength measurements from SOFIA/FORCAST and Spitzer. The symbols are the same for the other SEDs.}  
\label{C1}
\end{figure}

\begin{deluxetable*}{llccclll}[htbp]   
\tabletypesize{\footnotesize}
\tablenum{8}
\tablecaption{Model Results for RSGC1 Red Supergiants\tablenotemark{a}}
\tablewidth{0pt}
\tablehead{
\colhead{Star} & 
\colhead{Sp Type/Temp} & 
\colhead{Power law(r $^{n}$)} & 
\colhead{$\tau_{v}$} &
\colhead{r$_{1}$ (AU)} & 
\colhead{L$_{\odot}$} &
\colhead{M$_{\odot}$ yr$^{-1}$} & 
\colhead{Comment} 
}
\startdata 
	RSGC1-01*  & M3/M5/3550 & 1.6 & 0.5 & 83 & 3.35 $\pm$ 1.6 $\times$ 10$^{5}$ & 1.7 $\pm$ 0.5 $\times$ 10$^{-5}$  & SiO maser \\
	RSGC1-02*  & M4/M2/3700 & 1.7 & 0.5 & 69 & 2.15 $\pm$ 1.0 $\times$ 10$^{5}$ & 9.0 $\pm$ 2.5 $\times$ 10$^{-6}$ & SiO maser  \\
	RSGC1-03*   & M4/M5/3500  & 1.5  & 0.5 & 48 & 1.2 $\pm$ 0.5 $\times$ 10$^{5}$ & 1.5 $\pm$ 0.4 $\times$ 10$^{-5}$  &  \nodata \\
	RSGC1-04* & M0/M1/3800 & 1.7 & 0.4 &  93 & 3.8 $\pm$ 1.8 $\times$ 10$^{5}$ & 9.7 $\pm$ 2.8 $\times$ 10$^{-6}$ & SiO maser \\
RSGC1-05*  & M6/M4/3500 & 1.3 & 0.1 & 61 & 1.9 $\pm$ 0.9 $\times$ 10$^{5}$ & 8.2 $\pm$ 2.3 $\times$ 10$^{-6}$ & \\
RSGC1-06*  & M5/3400   & 1.9 & 0.5 & 68  &  2.3 $\pm$ 1.1 $\times$ 10$^{5}$ & 3.4 $\pm$ 1.0 $\times$ 10$^{-6}$ & \nodata  \\
RSGC1-07*  & M2/M3/3800 & 1.7 & 0.2 & 65 & 1.9 $\pm$ 0.9  $\times$ 10$^{5}$ & 3.4 $\pm$ 0.9 $\times$ 10$^{-6}$ & \nodata \\
RSGC1-08*   & M3/M3/3600 & 1.6  & 0.1 & 63 & 2.0 $\pm$ 0.9 $\times$ 10$^{5}$ & 2.6 $\pm$ 0.7 $\times$ 10$^{-6}$ & \nodata \\
RSGC1-09*   & M3/M6/3600 & 1.5  & 0.1  & 54 & 1.5 $\pm$ 0.7 $\times$ 10$^{5}$ & 3.5 $\pm$ 1.0 $\times$ 10$^{-6}$ & \\
RSGC1-10*  & M5/M3/3600 & 1.5 & 0.1 & 68  & 2.35 $\pm$ 1.1 $\times$ 10$^{5}$ & 4.4 $\pm$ 1.2 $\times$ 10$^{-6}$ & \\
RSGC1-11*   & M1/M4/3800  & 1.8 & 0.1  & 65 & 2.0 $\pm$ 0.9 $\times$ 10$^{5}$ & 1.7 $\pm$ 0.5 $\times$ 10$^{-6}$ & \nodata   \\ 
	RSGC1-12*   & M0/3900 & 2.0  & 0.1 & 64 & 1.9 $\pm$ 0.9 $\times$ 10$^{5}$ & 3.4 $\pm$ 1.0 $\times$ 10$^{-7}$ & \\ 
	RSGC1-13*   & M3/K2/4200   & 1.7  & 0.5  & 85  & 2.9 $\pm$ 1.4 $\times$ 10$^{5}$ &  2.7 $\pm$ 0.8 $\times$ 10$^{-5}$  &  SiO, H$_{2}$O, OH maser  \\ 
	RSGC1-14    &  M3/M1/3700      & 2     &  0.05    & 39    & 7.4  $\pm$ 3.5 $\times$ 10$^{4}$    &  1.9 $\pm$ 0.5 $\times$ 10$^{-7}$    & \\
	RSGC1-15*    &  G0/G6/\nodata   & \nodata   & \nodata & \nodata    & 6.2 $\pm$ 2.9 $\times$ 10$^{5}$:  \nodata   &  post RSG, see text\\
\enddata
\tablenotetext{a}{Stars with  FORCAST fluxes have an asterisk.}
\end{deluxetable*}

Maser emission is measured in four of the most luminous red supergiants; RSGC1-01, 02, and 04 show SiO emission while RSGC1-13 is the source of SiO, H$_{2}$O and OH emssion. All four are quite dusty and have relatively high mass loss rates.  The SED for RSGC1-13 in Figure 11  illustrates its strong silicate emission and very dusty excess emission out to 40~$\mu$m. Given its multiple maser emissions and earlier K spectral type, RSGC1-13 is probably the   most evolved RSG and has already begun its transition back to warmer temperatures. This possibility is supported by the presence of RSGC1-15, a probable yellow hypergiant member of the cluster. RSGC1-13 has the highest measured $\dot{M}$ in RSGC1, however it does not share the very high circumstellar extinction visible in the SED for RSGC2-49. This suggest that it has not yet experienced the high mass loss episodes visible in the ejecta and SEDs of many post-RSG hypergiants.   

\begin{figure}[h]   
\epsscale{0.7}
\plotone{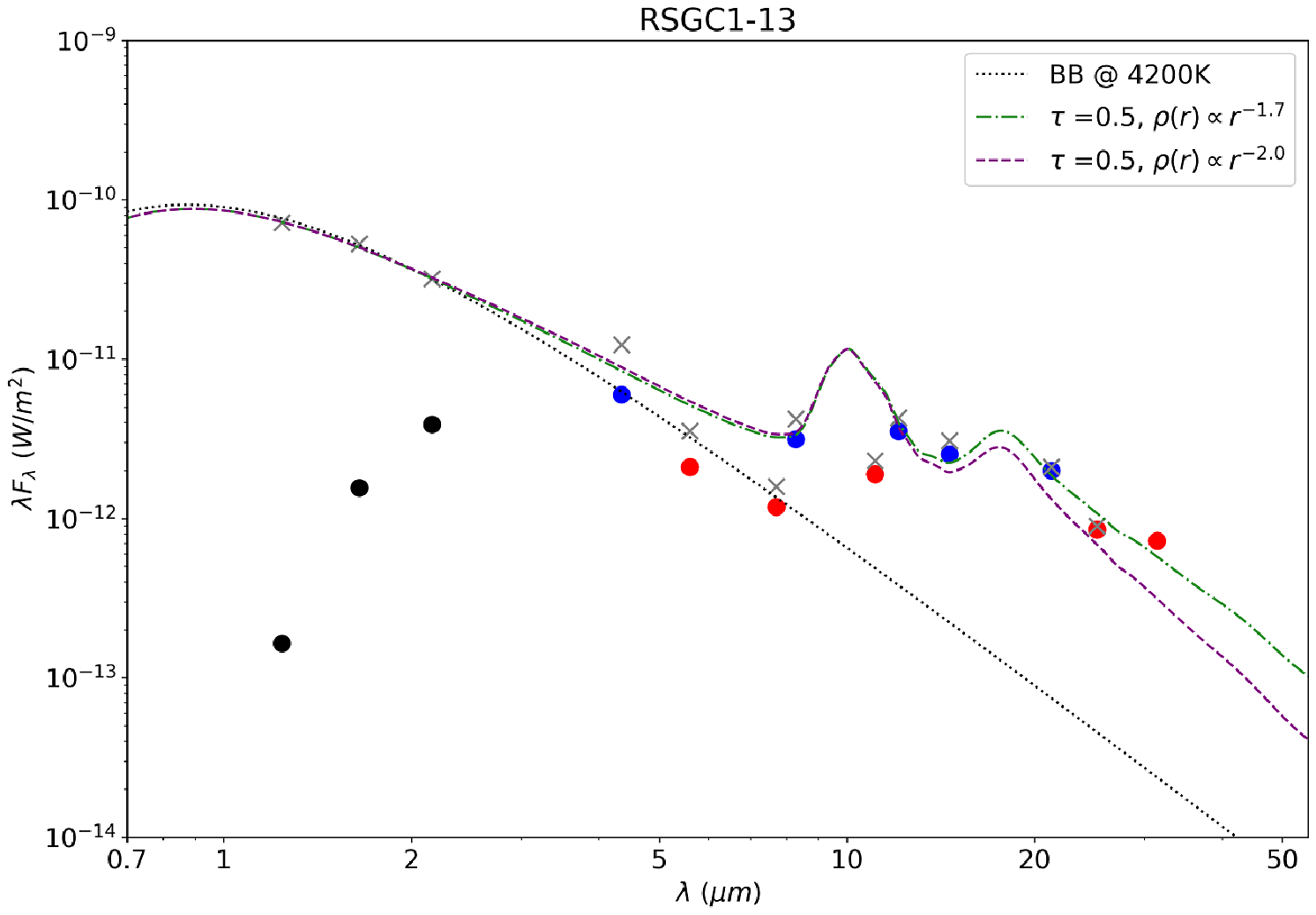}
\caption{The SED for the maser source RSGC1-13. Symbols are the same as in the other SEDs.}  
\end{figure}

Unfortunately, insufficient data is available for the yellow hypergiant RSGC1-15 at wavelengths beyond 7~$\mu$m to model its circumstellar dust  and $\dot{M}$. In addition to 2MASS data, published fluxes from Spitzer/IRAC and two data points  from the FORCAST images are available, but RSGC1-15 was not detected in the FORCAST images at 11~$\mu$m and longer.  The published spectral types \citep{Figer,Davies08} indicate a range from G0 to G6 based on the strength of the CO band heads, and spectra separated by only four months shows variable emission in the CO band heads \citep{Davies08}.  We get a temperature of $\approx$ 8000~K from the 2MASS fluxes alone and a corresponding luminosity of 6.2 $\times 10^{5} L_{\odot}$ compared to 6900~K and 2.3 $\times 10^{5} L_{\odot}$ from \citet{Davies08}.  This range in apparent temperature is not surprising for the hypergiants which often have optically-thick winds. RSGC1-15's variablity, possible  temperature range, and lack of measurable circumstellar dust is similar to the hypergiant $\rho$ Cas \citep{Shenoy16}.          

We have eight stars in common with the data for the RSGC1 red supergiants in \citet{BD2020}. Comparison shows similar derived luminosities well within the  quoted errors in both studies, but our mass lass rates estimated from the DUSTY model fits are significantly higher. \citet{BD2020} assume a constant mass loss rate in their DUSTY models but in this paper we have allowed the power law to vary. As we show for the RSGC1 supergiants as well as for red supergiants in the other clusters, a variable mass loss rate often yields a better representation of the long wavelength fluxes. 
This alternative result will affect the mass-loss rate-luminosity relation for the res supergiants discussed in the next section.

\section{Results and Discussion}

\subsection{The  Mass-Loss Rate -- Luminosity Relation for Red Supergiants}

The empirical $\dot{M}$ - Luminosity relation for red supergiants is critical to the models and evolutionary tracks for evolved stars. Not only does it allow the calculation of the total mass lost during this end stage for many stars, but
whether the rate has been constant or episodic may determine whether the star 
will evolve on a blue loop back to warmer temperatures before the terminal 
state. This latter possibility is now important with respect to the progenitors  of supernovae and possible alternative final stages for  massive stars.

Several $\dot{M}$ - L relations are available in the literature; including for example, \citet{Reimers}, \citet{vanloon} and by de~Jager and collaborators \citep{deJ,NJ90}. The de~Jager prescriptions   are the most commonly used in the evolutionary models. There are not only differences in the predictions among these relations, but also significant scatter within each. This latter should not be surprising though since it is reasonable to expect the stars to continue to
evolve during the RSG stage with $\dot{M}$ increasing with time, the $\dot{M}$ may be variable, and they may experience high mass loss episodes as observed in some of the  most evolved red supergiants.  \citet{Joss} have published a thorough review of the different published relations and conclude that the de~Jager prescriptions provide good representations of the expected mass loss rates for the Galactic RSGs with Solar metallicity.   

Use of clusters with numerous RSGs may reduce some of the observed internal scatter in the relation by removing some of the uncertainty due different distances and ages. In a series of papers \citet{BD2016,BD2018,BD2020}  have examined the  $\dot{M}$ - L relation for several clusters; NGC~7419, $\chi$~Per, NGC~2100 in the LMC, and RSGC1. RSGs with luminosities generally  below 10$^{5} L_{\odot}$ are represented by the stars in NGC 7419, $\chi$~Per, and NGC~2100, while the more luminous RSGs ($\geq$ 10$^{5} L_{\odot}$) are in  RSGC1.  The first group of clusters yields a $\dot{M}$ - L relation that is much steeper  than all of the previous prescriptions with several of the lower luminosity RSGs significantly below the 
\citet{deJ} and \citet{NJ90} relations. The more luminous RSGs in RSGC1 however lie on a distribution that is curiously offset from and even below that for the less luminous supergiants (see Figure~2 in \citealt{BD2020}).  The authors argue that  
this is an initial mass effect such that the more massive stars have a lower mass loss rate for a given luminosity. 

In this paper, we have treated the DUSTY models somewhat differently by allowing the mass loss rate to vary.  Our data include stars in NGC 7419 and RSGC1 in common with Beasor and Davies plus RSGC2 and RSGC3  with many stars in the  10$^{4}-10^{5}$ L$_{\odot}$ range.  

Our $\dot{M}$ - L relation for the four clusters is shown in three panels in 
Figure 12. The top panel (A) simply shows our data with the average  error in 
the logs for each cluster. In the middle panel (B) we've added the two linear relations from \citet{BD2020} for comparison and the bottom panel (C) is our data with the de~Jager prescriptions. In Figure 12C, we've added  some  high luminosity, high $\dot{M}$ evolved supergiants in the Solar neighborhood for comparison with the position of RSGC2-49, and also to illustrate these extreme stars not represented by the  general prescriptions. The mass loss rates for these stars are from our two previous papers using SOFIA/FORCAST data with DUSTY models. We adopt the same mass loss equation (A8) with a gas to dust ratio of 200 used for the RSG clusters in this paper for comparison.

\begin{figure}[h!]   
\figurenum{12}
\epsscale{0.6}
\plotone{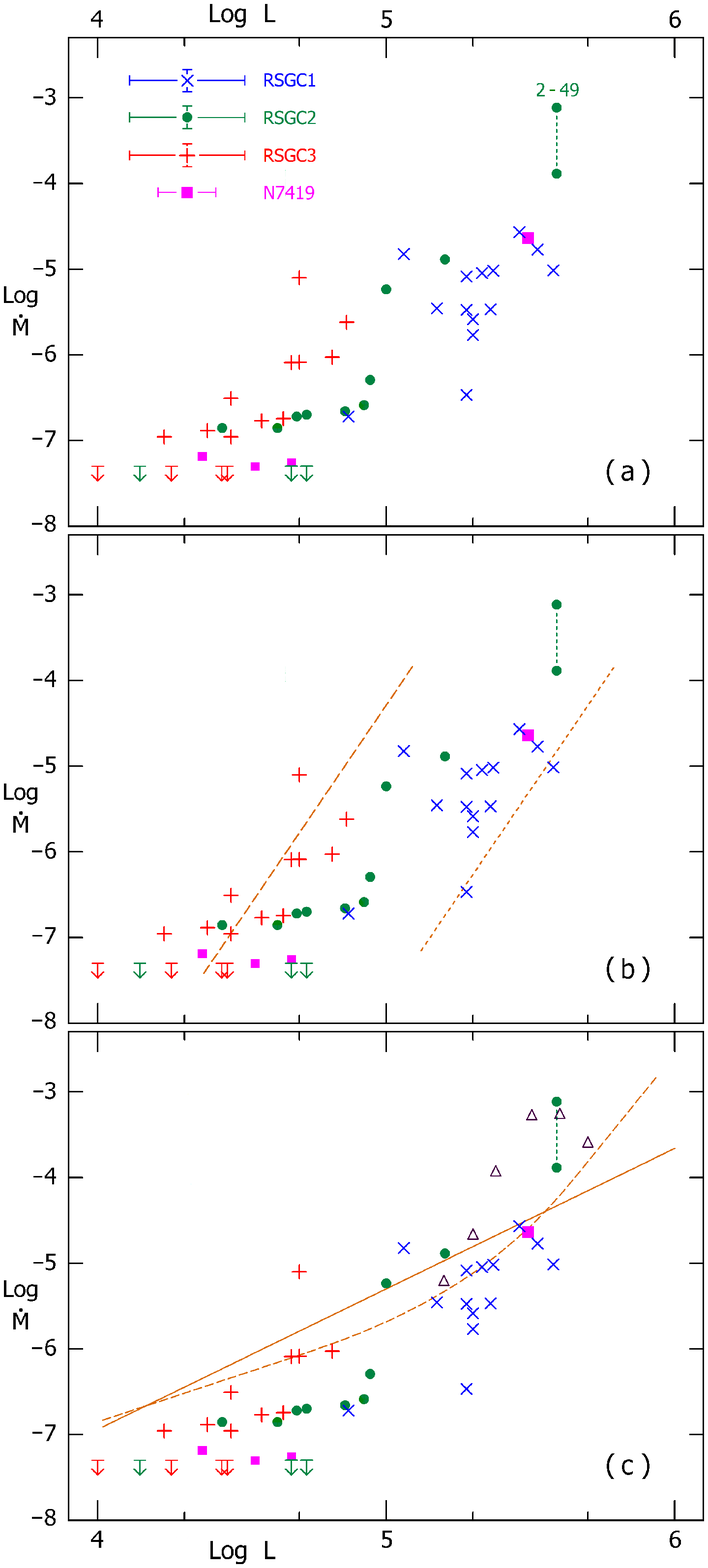}
\caption{Top(12A): The $\dot{M}$ - Luminosity relation for the RSGS in the four clusters. The color-coded identification for the different clusters is given with the errors in the logs. Those few stars with no infrared excess and no measurable mass loss rates are indicated by downward errors. Middle(12B): $\dot{M}$ - Luminosity relation for the RSGS in the four clusters with the linear relations from \citet{BD2020} shown as dashed lines in red. Bottom(12C): $\dot{M}$ - Luminosity relation for the RSGS in the four clusters with the prescriptions from \citet{deJ} and \citet{NJ90} for T$_{eff}$ 3750K. Several high luminosity, high $\dot{M}$ are also shown  as open triangles. Moving from left to right, with increasing luminosity, the plotted stars are $\mu$ Cep, S Per, VX Sgr, NML Cyg, VY CMa, IRC~10420. Note the position of $\mu$~Cep often considered to have an anomalously low $\dot{M}$, but with a gas to dust ratio of 200 it falls with the RSGC1 stars.}  
\end{figure}

 At luminosities below $\approx  10^{5} L_{\odot}$  our data show a significant population of RSGs with mass-loss rates that also lie below the  de~Jager et al. relations plus a few stars closer to the de~Jager curve. These latter stars in RSGC3 and RSGC2 have SEDS that show the presence of circumstellar dust and most are 
 fit with a steady constant mass loss rate.  We do not, however, confirm the steep slope for the lower luminosity RSGs reported by \citet{BD2018,BD2020}. We note 
 that their steep linear  relation is heavily weighted by stars in NGC~2100 in the LMC with a gas/dust ratio of 500. A lower ratio would reduce their $\dot{M}$ and yield a flatter relation. A higher v$_{exp}$ for the cluster RSGs would yield a higher $\dot{M}$. Velocities of 35- 40 km s$^{-1}$ are measured for some high luminosity RSGs, but these are are all $\dot{M}$ stars with complex circumstellar ejecta. We doubt that a higher v$_{exp}$ is applicable for these lower luminosity RSGs many of which have virtually zero mass-loss rates. Likewise, a higher gas to dust ratio will not move them substantially closer to the de~Jager curve. 
 
Our results could be described as representing a lower bound to the mass loss rates for these less luminous RSGs.   In their review, \citet{Joss} also find several RSGS in the Solar region with luminosities between log L$_{\odot}$ of  4.5 to 5.0 that lie below the deJager curve. Although their data do not extend below log  L$_{\odot}$ of 4.5, the distribution of their stars are consistent with the less luminous RSGs in Figure 12A.  

At luminosities above 10$^{5} L_{\odot}$ there is a  rapid transition to higher mass loss rates. Most of the stars, in RSGC1, are close to or just below the de~Jager et al.\ prescriptions in Figure 12C. This is in contrast to the much lower linear relation for the RSGC1 stars in \citet{BD2020}. These differences are obviously due to the treatment of results from the DUSTY models with a variable power law for the density distribution.  If we had used a slightly higher v$_{exp}$ (35--40 km s$^{-1}$), their $\dot{M}$ would move them closer to the de~Jager relation as would a lower T$_{in}$. Thus the distribution of the more luminous RSGs in our study is a reasonable approximation to the prescriptions determined empirically for RSGS in the Solar neighborhood. For example, the $\dot{M}$--L$_{\odot}$ relation in \citet{Joss} likewise shows the same shift to significantly higher mass loss rates for stars above 10$^{5} L_{\odot}$ in their study of 39 RSGs in the Solar neighborhood with an independent mass loss rate determination.  

The majority of stars used by \citet{deJ} in the Solar region are known semi-regular (SR) variables. The high $\dot{M}$ stars in Figure 12C are all known variables.  The role of variability on the mass loss rates and mass loss histories of the RSGs and other massive stars has been not fully explored. It seems reasonable to us that as stars evolve during  the red supergiant stage, their expanded envelope becomes more subject to instabilities that thus enhance their mass loss rate. 

Most stellar models and evolutionary tracks have relied on the de~Jager et al.\ $\dot{M}$ - Luminosity relations for the red supergiants. Our results for the RSGs below $\approx 10^{5} L_{\odot}$ suggests that a lower $\dot{M}$  should be adopted as an alternative or better, a  possible range of mass loss rates for the less luminous  RSGs.  Instead of a single linear or curved relation, the $\dot{M}$ - Luminosity relation may be better represented by a broad band, approximately parallel to the de~Jager curve, with a lower bound defined by the  low luminosity  RSGs in RSGC3 and RSGC2 which at 10$^{5} L_{\odot}$ curves more rapidly upward.  This is speculative, but the  transition to much higher $\dot{M}$ at about 10$^{5} L_{\odot}$ corresponds approxmately to an initial mass of  18--20 M$_{\odot}$ which is interestingly close to the upper limit for RSGs becoming Type II SNe.

\subsection{Evolution in the Red Supergiant Stage}  

The large number of red supergiants in these four clusters provide a sample with the expected range in RSG luminosities and initial masses from $\approx$ 9 to $\geq$ 25 M$_{\odot}$. They also include examples of more evolved RSGs, such as MY Cep and RSGC2-02, several sources of maser emission, and candidates for post-red supergiant evolution well-represented by RSGC2-49. 

Several questions about red supergiant evolution concern their eventual fate 
as the progenitors of Type II-P supernovae, or possible evolution back to warmer temperatures and their eventual demise as alternative supernovae or in a direct collapse to a black hole. These alternate possibilities very likely depend on
their mass loss histories. In the previous section, we showed a complex dependence of $\dot{M}$ on luminosity with a strong shift to higher mass loss rates as the luminosities of the stars increase above $10^{5} L_{\odot}$. 

Another question concerns how the stars evolve during the red supergiant stage. Do they slowly ascend the RSG region much like red giants getting more luminous at essentially the same temperature as suggested by \citet{Davies13}, or do 
they evolve through the RSG stage slowly getting cooler with later spectral types and 
increased $\dot{M}$? The former is based on a near-constant temperature for RSGs where the TiO based spectral type is a luminosity indicator. The latter model does not rule out some increase in luminosity as the envelope expands and cools.

The HR Diagrams for the four clusters are shown separately in Figure 13 using the luminosities and  temperatures in Tables 5 -- 8. 

To examine evolution in the red supergiant stage, we first look at the luminosity 
dependence on the apparent temperature of the stars in each cluster, and as a second test, we examine the temperature in different luminosity ranges. Since there is more than one published spectral type for most stars, even when the authors use the same data, as in RSGC1, we adopt the temperature from our SED fits. This is not strictly a test of the Davies et al hypothesis in which the RSGs all have essentially the same temperature. The results are shown in Table 9 binned by temperature and by luminosity for the three RSG clusters. With only four stars, all about the same temperature and luminosity, N7419 is not included, and in the subsequent discussion we show that MY~Cep has very likely evolved from a more massive star.

\begin{splitdeluxetable*}{lccllBlclcc}  
\tabletypesize{\footnotesize}
\tablenum{9}
\tablecaption{Luminosity--Temperature Dependence}
\tablewidth{0pt}
\tablehead{
\colhead{Cluster} &
\colhead{N(RSGs)} &
\colhead{Temp Range (K)} & 
\colhead{Lum Range (L$_{\odot}$)} & 
\colhead{Mean Lum  (L$_{\odot}$)} &
\colhead{Cluster} &
\colhead{N(RSGs)} &
\colhead{Lum Range (L$_{\odot}$)} &
\colhead{Temp Range (K)} &
\colhead{Mean Temp (K)} 
} 
\startdata 
Binned by Temperature &    &    &    &  & Binned by Luminosity  &    &    &    &   \\
RSGC3 & 9  &  3800--4000  & 1.1--4.7 $\times 10^{4}$ & 2.6 $\pm$ 0.3 $\times 10^{4}$ &
RSGC3  &  12   &  $10^{4}$-- $< 5 \times 10^{4}$ &  4000--3200 &  3740 $\pm$ 70 \\
      & 4  &  3500--3600  & 3.7--7.3 $\times 10^{4}$ & 5.6 $\pm$ 0.7 $\times 10^{4}$ & 
      &  4    &  5 $\times 10^{4}$ -- $< 10^{5}$  & 3600--3200 &  3675 $\pm$ 80  \\ 
      & 3  &  3200--3300  & 2.4--5.0 $\times 10^{4}$ & 3.9 $\pm$ 0.6 $\times 10^{4}$ &   
      & \nodata  & \nodata  &\nodata   & \nodata    \\  
RSGC2 & 5  & 3700-4000 & 1.4--8.4  $\times 10^{4}$ &  4.8 $\pm$ 1.0 $\times 10^{4}$ & 
RSGC2  &  5    &  $10^{4}$ -- $< 5 \times 10^{4}$ &  4000--3600 &  3720 $\pm$ 65 \\
      &  6 & 3400--3600 & 2.7--10 $\times 10^{4}$ & 6.5 $\pm$ 1.0 $\times 10^{4}$  &  
      &  5    &  5 $\times 10^{4}$ -- $< 10^{5}$  & 3900--3400 &  3640 $\pm$ 80  \\
      &  1 &  3200      & 1.6 $\times 10^{5}$ &   \nodata  &  
      &  2    &  $10^{5}$ --5 $\times 10^{5}$ &   3400--3200  &  3300  \\ 
RSGC1 &  6  & 3700--3900  & 0.7--3.8 $\times 10^{5}$  &  2.1 $\pm$ 0.9 $\times 10^{5}$ &  
RSGC1  &  1    &  5 $\times 10^{4}$ -- $< 10^{5}$ &  3700       &  \nodata \\
      &  6  &  3400--3600 & 1.2--3.4 $\times 10^{5}$  &  2.0 $\pm$ 0.7 $\times 10^{5}$ &  
      &  12   &  $10^{5}$-- 5 $\times 10^{5}$ &   3900--3400 &  3650 $\pm$ 40  \\  
\enddata
\end{splitdeluxetable*}

The results for RSGC1 and RSGC3, binned by temperature, do not support a luminosity dependence on the apparent 
surface temperature. The luminosity ranges for the different temperature bins overlap. The luminosity distribution of the stars in RSGC2 show some dependence 
on temperature but the signature is not strong and not statistically significant. The second test, binned by luminosity, supports this conclusion. The range of temperatures in each luminosity range is esentially the same.  
The highest luminosity range for RSGC2, with only two stars and  lower temperatures, supports evolution in the RSG stage to cooler, later types since these are the most luminous, and presumably most massive, evolved members.  We conclude that there is no strong evidence that red supergiants evolve significantly up the RSG branch.

The HRDS  for the four clusters clearly illustrate the  wide range in luminosities, initial masses, and ages of the stars that define the RSG range in the HR 
Diagram. Evolutionary tracks from \citet{Ekstrom}, without rotation,  are shown on each HRD with initial masses selected to correspond to the range of luminosities of the red supergiants in each cluster. Since NGC 7419 has a large number of Be star members which are rapid rotators, we also show the corresponding tracks with rotation for comparison. In the following discussion, we also compare the HRDs with the isochrones from \citet{Ekstrom}. 

\begin{figure}[h!]   
\figurenum{13}
\epsscale{1.0}
\plotone{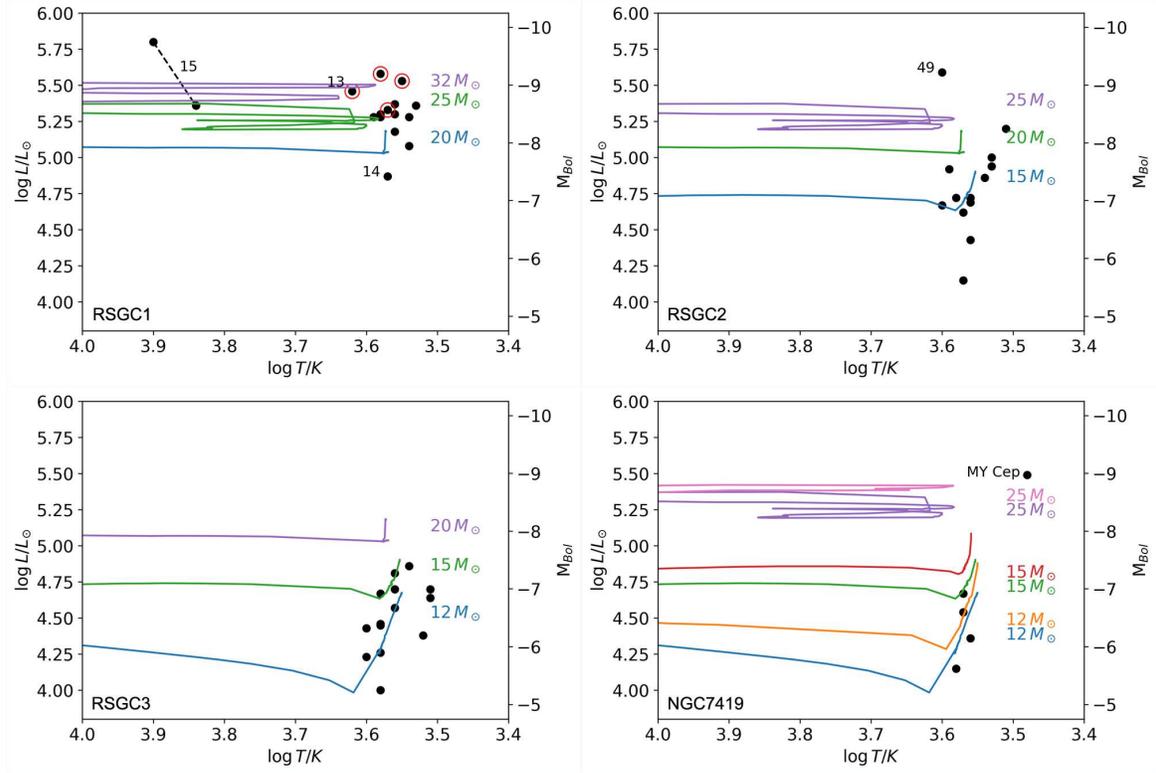}
	\caption{The HR Diagrams for the four clusters with the evolutionary tracks from \citet{Ekstrom} without rotation. The upper curve in the pairs in the  HRD for NGC 7419 is the corresponding model with rotation.}  
\end{figure}

The red supergiants in RSGC1 are mostly clumped in a relatively small luminosity range from $\approx$ $10^{5}$ to 5 $\times 10^{5}$ L$_{\odot}$ implying an initial mass range of 20 to 30 M$_{\odot}$.The post-RSG candidate, star 15, may have arisen from a higher mass star but the models for the higher mass ranges show multiple transits across the HRD. We also can't rule out the possibility that the three most luminous members have been on a post-RSG transit. As already mentioned, star 13 is candidate for beginning its evolution back to higher temperatures. 
Except for star 14, the red supergiants in RSGC1 are consistent with an age no more than 8 Myr. Star 14 is the star discussed earlier with little no circumstellar 
dust  selected by \citet{BD2020}  to estimate the foreground extinction. We likewise adopted its extinction. We note that its SED, lack of dust, and very low $\dot{M}$ are quite different from the other members. It is also located more than 2 $\arcmin$ from the center of the cluster. We therefore suggest that it may not be a member.

The red supergiant members of RSGC3 show a wide range in luminosities from $10^{4}$ to almost $10^{5}$ indicating a range in initial masses from $\approx$ 9  to somewhat above 15 M$_{\odot}$. Despite this wide range in luminosities and initial masses, the isochrones show that the position of these RSGs on the HRD are consistent with a common age of $\approx$ 16 Myr.  The lower luminosity stars may have recently entered the RSG stage while the most luminous with initial 
masses of $\sim$ 15 M$_{\odot}$ are nearing the end of their red supergiant lives.

RSGC2 and NGC 7419 are the exceptions to a common age for the apparent member red supergiants. RSGC2 has a large range of luminosities overlapping with both RSGC3 and RSGC1, and implying an initial mass range from 12 M$_{\odot}$ to more than 25 M$_{\odot}$ for the post-RSG star 49. As already noted in our previous discussion, the candidate members are spread over more than 6 $\arcmin$. This is also a very complex region. \citet{Negueruela12}  emphasized the extent of this region and concluded that RSGC2 is not an isolated cluster but part of  a region with numerous massive stars. We therefore suggest that RSGC2 and the surrounding red supergiants is more like a stellar association with populations of massive stars not all the same age. The stars may range in age from 16 Myr to less than 8 Myr. 

Comparison of the  isochrones with the evolutionary tracks for NGC 7419 suggests  that the four  lower luminosity red supergiants have essentially the same age, between 16 and 20 Myr with a relatively small range in initial mass from 10 to 15 M$_{\odot}$.  N7419 has a significant population of normal B-type stars as well as Be stars. \citet{Negueruela13} noted that the most luminous B stars, on the main sequence have  M$_{V}$'s of $\approx$ -5 mag. Assuming their early B2 spectral types, their bolometric luminosities would be -6.2 with corresponding 
initial masses of about 12 M$_{\odot}$. They and their immediate slightly more massive counterparts are the likely progenitors of the four red supergiants.  

MY Cep however is not consistent with their inferred age and initial masses.  
There is a large gap in luminosity between MY Cep and the four fainter RSGs. MY Cep is easily 6 to 10 times more luminous. Thus it is most likely evolved from a somewhat more massive star. If we assume that it began as a 20 M$_{\odot}$ star, based on the models it would be at most $\approx$ 9 Myr old.  The models for even more massive stars show multiple transits across the HRD, but they have similar maximum ages  at the end of the tracks.  Our previous discussion cast doubt on significant  evolution up the RSG branch  accompanied by a large increase in luminosity. But, if so, one would then have to ask  where are the missing red  supergiants that shoud be found on the HRD  between MY Cep and the four fainter stars.  An alternative explanation is that MY Cep is not a member. NGC 7419 is in the Perseus spiral arm and in its direction there are a several large associations with mixed populations of massive and evolved stars. 

\section{Final Remarks and Recommendations}  

The numerous red supergiants in these four clusters have provided us with a large sample covering the  expected initial masses and luminosities for the RSGs. Thus in this paper we have covered a wide range of topics from their SEDs, mass loss rates, the HRDs and their evolutionary state. We conclude that there is no strong statistical support for evolution up the RSG branch, but instead we support evolution  during the RSG stage to lower temperatures as their outer envelopes expand with increased mass loss.      

Our fits to the SEDs with the DUSTY models allow for a variable power law for the density distribution function and our estimate of the mass loss rates. Our $\dot{M}$ -- Luminosity relation (Figure 12) for 42 RSGs in the clusters  follows the \citet{deJ} relation but at luminosities below $10^{5} \;  L_{\odot}$, we find a  
population of RSGs with significantly lower $\dot{M}$. At higher luminosities, there is a rapid transition to higher mass loss rates. We note that \citet{Joss} show a  similar distribution of RSGS in the $\dot{M}$ --$L$ plane for 39 red supergiants in the Solar neighborhood based on an independent $\dot{M}$ from the flux at 60$\mu$m. Based on our results, we recommend that a lower mass loss rate be adopted, on average, for RSGs below $10^{5} \;  L_{\odot}$.  Instead of a single linear or curved relation, we suggest that a curved band, parallel to the de Jager relation, is a better representation of the  empirical relation with a  lower boundary represented by the less luminous RSGs which then curves upward at $10^{5} \;  L_{\odot}$.   For application to stellar models, the adopted  $\dot{M}$ corresponding to the luminosity could slowly increase as the star evolves during the RSG phase. 

\acknowledgments

We thank Kris Davidson for useful comments and critical reading of the paper.
We also thank the LBTI team for making the LMIRCam available for our project and for their support with planning and executing the observations, and the SOFIA Science and Mission Operations team for observation planning and the  
data reduction. This work was supported by NASA through award SOF 06-0089 to R. M. Humphreys isssued by USRA.

\vspace{5mm}
\facilities{SOFIA(FORCAST), LBT/LBTI(LMIRCam) }


\appendix 
\section{The Mass-Loss Rate from DUSTY}

DUSTY calculates the optical depth through a model dust shell down to the star using the column number density of dust grains that follow a particular grain size distribution. This size distribution, $n(a)$, is defined such that $n(a)da$ is the number of dust grains per cm$^2$ within the size interval $da$ along the line of sight. Using Mie theory, we can define the effective extinction efficiency $Q_{\rm{eff}}$ at some wavelength as

\begin{eqnarray}
{Q_{\rm{eff}}} = \frac{{\int_{{a_{\min }}}^{{a_{\max }}} {Q(a)n(a)\pi {a^2}da} } }{{\int_{{a_{\min }}}^{{a_{\max }}} {n(a)\pi {a^2}da} }}
\end{eqnarray}

The optical depth can be expressed as

\begin{eqnarray}
{\tau} = {Q_{\rm{eff}}}\int_{{a_{\min }}}^{{a_{\max }}} {n\left( a \right)\pi {a^2}da}
\end{eqnarray}

Using a dust grain mass density $\rho_{\rm{d}}$, the mass column depth corresponding to the optical depth is

\begin{eqnarray}
m\left( {{\rm{gm}}\;{\rm{cm}}{}^{ - 2}} \right) = \int_{{a_{\min }}}^{{a_{\max }}} {\rho _d}{n\left( a \right)} \frac{4}{3}\pi {a^3}da
\end{eqnarray}

Using an MRN grain size distribution ($n_0$ in $\rm{cm}^{-3}$) 

\begin{eqnarray}
n(a) = {n_0}{\left( {\frac{{{a_0}}}{a}} \right)^{3.5}}
\end{eqnarray}

and after some algebra we have

\begin{eqnarray}
m = \sqrt {{a_{\max }}{a_{\min }}} \frac{4}{3}\frac{{{\rho _d}\tau }}{{{Q_{\rm{eff}}}}}
\end{eqnarray}

Although DUSTY softens the dust grain size distribution at $a_{\max}$, this has only a
minor effect on our simplified formula. DUSTY uses a radial density distribution that
 follows a power law with index $n$. Using this formulation we have

\begin{eqnarray}
\rho (r) &=& {\rho _{{r_{\min }}}}{\left( {\frac{{{r_{\min }}}}{r}} \right)^n} \nonumber \\
  m &=& \int_{{r_{\min }}}^{{r_{\max }}} {\rho (r)} dr \sim \frac{{{\rho _{{r_{\min }}}}
  {r_{\min }}}}{{n - 1}}\quad {\rm{for }}~n > 1~{\rm{ and }}~{r_{\max }} >  > {r_{\min } } 
\end{eqnarray}

Equating the column mass in eqs. 5 and 6 and solving for $\rho_{\min}$, we have

\begin{eqnarray}
{\rho _{{r_{\min }}}} = \frac{{m\left( {n - 1} \right)}}{{{r_{\min }}}} = \frac{{\sqrt {{a_{\max }}{a_{\min }}} }}{{{r_{\min }}}}\frac{{4{\rho _d}\left( {n - 1} \right)}}{3}\frac{\tau }{{{Q_{\rm{eff}}}}}
\end{eqnarray}

Following \citet{Shenoy16} we can compute the average mass-loss rate over the lifetime for the model shell by assuming a constant wind velocity $V$ and integrating the density from $r_{\min}$ to $r_{\max}$ as follows

\begin{eqnarray} \label{eq:A8}
{M_{tot}} &=& \int_{{r_{\min }}}^{{r_{\max }}} {4\pi {r^2}} \rho \left( r \right)dr = \frac{{4\pi {\rho _{{r_{\min }}}}}}{{3 - n}}\left( {r_{\max }^{3 - n}r_{\min }^n - r_{\min }^3} \right) \nonumber \\
\dot M &=& \frac{{{M_{tot}}V}}{{{r_{\max }} - {r_{\min }}}} \sim \frac{{16\pi }}{3}\frac{{n - 1}}{{3 - n}}\sqrt {{a_{\max }}{a_{\min }}} {\rho _d}\frac{\tau }{{{Q_{\rm{eff} }}}}r_{\max }^{2 - n}r_{\min }^{n - 1}v_{exp}
\end{eqnarray}

where we assume $r_{\max} >> r_{\min}$. For the case where the index $n=2$, a constant mass-loss rate, we have

\begin{eqnarray}
\dot M \sim \frac{{16\pi }}{3}\sqrt {{a_{\max }}{a_{\min }}} {\rho _d}\frac{\tau }{{{Q_{\rm{eff}}}}}{r_{\min }}v_{exp}
\end{eqnarray}

DUSTY outputs the optical depth for the model dust shell at a wavelength of $0.55~\micron$, so $Q_{\rm{eff}}$ is calculated at this wavelength using the optical constants for cool O-rich silicates from \citep{Ossen}. Using optical constants $n=2.1$, $k=0.089$ and a range in grain size $a_{\min}=0.005~\micron$, $a_{\max}=0.25~\micron$ we compute $Q_{\rm{eff}}=0.38$ at V.

\section{SOFIA/FORCAST Fluxes}

\begin{deluxetable*}{cccc}[htbp]  
\tabletypesize{\footnotesize}
\tablewidth{0pt}
\tablecaption{New SOFIA/FORCAST Photometry for Stars in NGC7419 }
\tablehead{
\colhead{Star} & \colhead{11.1 $\mu$m (Jy)} & \colhead{31.5 $\mu$m (Jy)} & \colhead{37.1 $\mu$m (Jy)} 
 }
\startdata 
 NGC7419-B435 & 1.53 $\pm$ 0.25 & ... & ... \\
 NGC7419-B696 & 1.78 $\pm$ 0.85 & 0.35 $\pm$ 0.52\tablenotemark{a}& ... \\
 NGC7419-B921 & 1.24 $\pm$ 0.78 & ... & ... \\
 NGC7419-B950\tablenotemark{b} & 101.3 $\pm$ 5.6 & 33.3 $\pm$ 2.7 & 23.09 $\pm$ 0.81 \\
 & & 41.1 $\pm$ 2.9\tablenotemark{a}& \\
\enddata
\tablenotetext{a}{Level 4 data} 
\tablenotetext{b}{15\farcs36 aperture diameter}
\end{deluxetable*}

\begin{deluxetable*}{ccccccc}[htbp]
\tabletypesize{\footnotesize}
\tablewidth{0pt}
\tablecaption{New SOFIA/FORCAST Photometry for Stars in RSGC1 }
\tablehead{
\colhead{Star} & \colhead{5.6 $\mu$m} & \colhead{7.7 $\mu$m} & \colhead{11.1 $\mu$m} & \colhead{25.3 $\mu$m} & \colhead{31.5 $\mu$m} \\
   & \colhead{(Jy)} & \colhead{(Jy)} & \colhead{(Jy)} & \colhead{(Jy)} & \colhead{(Jy)} 
 }
 \startdata 
 RSGC1-01 & 6.20 $\pm$ 0.19 & 5.20 $\pm$ 0.19 & 13.76 $\pm$ 0.31 & 11.5 $\pm$ 3.2 & 9.8 $\pm$ 2.7  \\
 RSGC1-02 & 6.56 $\pm$ 0.32 & 5.65 $\pm$ 0.26 & 14.20 $\pm$ 0.01 & 13.1 $\pm$ 3.5 & 11.3 $\pm$ 3.8 \\
 RSGC1-03 & 4.38 $\pm$ 0.05 & 4.27 $\pm$ 0.11 & 8.1 $\pm$ 2.9 & 7.3 $\pm$ 1.1 & 6.35 $\pm$ 0.02 \\
 RSGC1-04 & ...\tablenotemark{a} & 2.43 $\pm$ 0.11 & ...\tablenotemark{a} & 5.58 $\pm$ 0.81 & 4.4 $\pm$ 1.1 \\
 RSGC1-05\tablenotemark{b} & 2.68 $\pm$ 0.03 & 2.76 $\pm$ 0.04 & 2.97 $\pm$ 0.76 & 2.2 $\pm$ 1.4 & 1.24 $\pm$ 0.02 \\
 RSGC1-06 & 2.69 $\pm$ 0.02 & 2.81 $\pm$ 0.01 & 2.9 $\pm$ 3.7 & 1.68 $\pm$ 0.52 & 1.45 $\pm$ 0.66 \\
 RSGC1-07 & 2.51 $\pm$ 0.002 & 2.40 $\pm$ 0.15 & 2.80 $\pm$ 0.18 & 1.23 $\pm$ 0.21 & 1.05 $\pm$ 0.36 \\
 RSGC1-08\tablenotemark{b} & 2.59 $\pm$ 0.03 & 2.16 $\pm$ 0.04 & 2.16 $\pm$ 0.76 & 1.7 $\pm$ 1.4 & 1.30 $\pm$ 0.02 \\
 RSGC1-09 & 2.46 $\pm$ 0.15 & 2.40 $\pm$ 0.07 & 3.5 $\pm$ 1.3 & 1.41 $\pm$ 0.56 & 1.36 $\pm$ 0.34 \\
 RSGC1-10 & 2.06 $\pm$ 0.08 & 2.07 $\pm$ 0.17 & 2.4 $\pm$ 3.2 & 1.53 $\pm$ 0.36 & 1.67 $\pm$ 0.71 \\
 RSGC1-11 & ...\tablenotemark{a} & 1.92 $\pm$ 0.05 & ...\tablenotemark{a} & ...\tablenotemark{a} & 1.14 $\pm$ 0.27 \\
 RSGC1-12 & 1.64 $\pm$ 0.07 & 1.61 $\pm$ 0.12 & 1.8 $\pm$ 1.4 & 1.18 $\pm$ 0.07 & 0.53 $\pm$ 0.91 \\
 RSGC1-13 & 3.92 $\pm$ 0.31 & 3.02 $\pm$ 0.01 & 7.04 $\pm$ 0.83 & 7.17 $\pm$ 0.68 & 7.6 $\pm$ 1.4 \\
	RSGC1-15 &  1.025 $\pm$ 0.002 &  0.465 $\pm$ 0.002  & \nodata & \nodata & \nodata \\  
\enddata
\tablenotetext{a}{Outside FOV} 
\tablenotetext{b}{RSGC1-05 and RSGC1-08 are merged. One 15.36'' aperture was fit around both stars to obtain a total flux. The ratios of their peaks were applied to this total value to determine approximate individual flux values.} 
\end{deluxetable*}

RSGC1 is the only cluster with published SOFIA/FORCAST fluxes from another source.  Comparison with the fluxes for the eight stars in common with  RSGC1 in \citet{BD2020} shows that our aperture-based photometry is systematically lower with respect to their PSF measurements. The differences vary from star to star and with wavelength.  They average $\approx$ 0.1 Jy for the two shortest wavelengths and up 0.5 --0.6 Jy $\geq$ 11$\mu$m.

\begin{deluxetable*}{ccccccc}[htbp] 
\tabletypesize{\footnotesize}
\tablewidth{0pt}
\tablecaption{New SOFIA/FORCAST Photometry for Stars in RSGC2 }
\tablehead{
\colhead{Star} & \colhead{7.7 $\mu$m} & \colhead{11.1 $\mu$m} & \colhead{19.7 $\mu$m} & \colhead{25.3 $\mu$m} & \colhead{31.5 $\mu$m} & \colhead{37.1 $\mu$m} \\
   & \colhead{(Jy)} & \colhead{(Jy)} & \colhead{(Jy)} & \colhead{(Jy)} & \colhead{(Jy)} & \colhead{(Jy)} 
 }
\startdata 
 RSGC2-02\tablenotemark{b} & 6.64 $\pm$ 0.40 & 20.1 $\pm$ 1.0 & 15.70 $\pm$ 0.12 & 11.34 $\pm$ 0.30 & 8.20 $\pm$ 0.06 & 5.79 $\pm$ 0.05 \\
 & & & & & & 5.7 $\pm$ 1.9 \\
 & & & & & & 5.77 $\pm$ 0.22\tablenotemark{a} \\
 RSGC2-03 & ...\tablenotemark{c} & ...\tablenotemark{c} & ...\tablenotemark{c} & 1.15 $\pm$ 0.71 & 0.95 $\pm$ 0.02 & 0.67 $\pm$ 0.52\\
 & & & & & & 0.69 $\pm$ 0.45\tablenotemark{a} \\
 RSGC2-06 & 1.25 $\pm$ 0.38 & 1.32 $\pm$ 0.43 & 0.66 $\pm$ 0.05 & 0.31 $\pm$ 0.17 & ... & ... \\
 RSGC2-08 & 0.62 $\pm$ 0.39 & 0.84 $\pm$ 0.08 & 0.3 $\pm$ 1.1 & 0.67 $\pm$ 0.33 & 0.57 $\pm$ 0.56 & ...\\
 RSGC2-10 & 0.84 $\pm$ 0.52 & ... & ... & ... & ... & ... \\
 RSGC2-11 & 1.24 $\pm$ 0.61 & 1.23 $\pm$ 0.99 & ... & ... & ... & ... \\
 RSGC2-14 & 0.58 $\pm$ 0.53 & ... & ... & ... & ... & ... \\
 RSGC2-15 & 0.68 $\pm$ 0.01 & ... & ... & ... & ... & ... \\
 RSGC2-18 & 0.28 $\pm$ 0.70 & ... & ... & ... & ... & ... \\
 RSGC2-52 & ...\tablenotemark{c} & ...\tablenotemark{c} & ...\tablenotemark{c} & ... & ... & 0.91 $\pm$ 0.20 \\
 & & & & & & 0.94 $\pm$ 0.16\tablenotemark{a}\\ 
\enddata
\tablenotetext{a}{Level 4 data} 
\tablenotetext{b}{15\farcs36 aperture diameter} 
\tablenotetext{c}{Outside FOV}
\end{deluxetable*}

\begin{deluxetable*}{cccccc}[htbp]  
\tabletypesize{\footnotesize}
\tablewidth{0pt}
\tablecaption{New SOFIA/FORCAST Photometry for Stars in RSGC3 }
\tablehead{
\colhead{Star} & \colhead{5.6 $\mu$m} & \colhead{7.7 $\mu$m} & \colhead{11.1 $\mu$m} & \colhead{25.3 $\mu$m} & \colhead{31.5 $\mu$m} \\
   & \colhead{(Jy)} & \colhead{(Jy)} & \colhead{(Jy)} & \colhead{(Jy)} & \colhead{(Jy)}
 }
\startdata 
 RSGC3-01 & 1.43 $\pm$ 0.30 & 0.95 $\pm$ 0.22 & ... & ... & ...\\
 RSGC3-02 & ...\tablenotemark{a} & 0.77 $\pm$ 0.24 & ... & ... & ...\\
 RSGC3-04 & ...\tablenotemark{a} & 1.70 $\pm$ 0.009 & ... & ... & ...\\
 RSGC3-05 & 1.65 $\pm$ 0.10 & 1.29 $\pm$ 0.14 & ... & ... & ...\\
 RSGC3-06 & 0.87 $\pm$ 0.0004 & 0.31 $\pm$ 0.002 & ... & ... & ...\\
 RSGC3-07 & 0.96 $\pm$ 0.61 & 0.59 $\pm$ 0.15 & 1.22 $\pm$ 0.56 & ... & ...\\
 RSGC3-10 & 0.73 $\pm$ 0.28 & 0.37 $\pm$ 0.11 & ... & ... & ...\\
 RSGC3-A11\tablenotemark{b} & 0.59 $\pm$ 0.32 & 0.35 $\pm$ 0.16 & ... & ... & ...\\
 RSGC3-27 & ...\tablenotemark{a} & 0.17 $\pm$ 0.20 & ... & ... & ...\\
\enddata
\tablenotetext{a}{Outside FOV}
\tablenotetext{b}{No cross-ID number in Clark et al. (2009)} 
\end{deluxetable*}

\end{document}